\documentclass[parskip, 12pt, times]{article}

\usepackage{graphics}
\usepackage{amssymb}
\usepackage{amsmath, framed}
\usepackage{braket}

\usepackage{epsfig}
\usepackage{amsmath}
\usepackage{amsfonts}
\usepackage{amssymb}
\usepackage{enumerate}

\usepackage{color,soul}

\usepackage{ifpdf}
\usepackage{	qcircuit}
\usepackage{pifont}
\usepackage{graphicx}
\usepackage{braket}
\usepackage{pifont}
\usepackage{array}
\usepackage{longtable}
\usepackage{bm}
\usepackage{hyperref}
\usepackage{color}
\usepackage{makeidx}
\usepackage{cite}
\usepackage{mathtools}
\usepackage{bbm}
\usepackage{wrapfig}
\usepackage[skip=0pt]{caption}
\usepackage{titlesec}
\usepackage[a4paper, margin=2cm, marginparwidth=0cm, marginparsep=0cm, outer=0cm, headheight=11pt, marginpar=0pt]{geometry}

\usepackage[font=footnotesize]{caption}

\usepackage[table]{xcolor}
\definecolor{matteRed}{RGB}{222,115,116} % Defining an Excel matte red color
\definecolor{matteYellow}{RGB}{255,176,59}
\definecolor{matteGreen}{RGB}{198,224,180} % Approximate Excel matte green color
\definecolor{matteBlue}{RGB}{141,180,226}  % Approximate Excel matte blue color

\usepackage{adjustbox}

\usepackage{tikz}

\newcommand{\Real}{\mathbb{R}}
\newcommand{\Complex}{\mathbb{C}}

\newcommand{\ketbra}[2]{| #1\rangle\!\langle #2 |}

\newcommand{\Tr}{\mathrm{Tr}}

\newcommand{\spana}{\text{span}}

\newcommand{\inv}{{\,\text{-}\hspace{-1pt}1}}
\newcommand{\ad}{\mathrm{ad}}
\newcommand{\Ad}{\mathrm{Ad}}

\newcommand{\g}{\mathfrak g}
\newcommand{\h}{\mathfrak h}

\renewcommand{\a}{\mathfrak a}
\renewcommand{\k}{\mathfrak k}
\newcommand{\p}{\mathfrak p}
\newcommand{\m}{\mathfrak m}

{\endMakeFramed}


\titleformat{\chapter}[display]
    {\normalfont\LARGE\bfseries}{\chaptertitlename\ \thechapter}{5pt}{\Large}
\titlespacing*{\chapter}{0pt}{-20pt}{5pt}
\titlespacing*{\section}{0pt}{8pt}{8pt}
\titlespacing*{\subsection}{0pt}{8pt}{8pt}

\DeclarePairedDelimiterX{\inp}[2]{\langle}{\rangle}{#1, #2}

\begin{document}

\newpage

\title{Solving the $KP$ problem with the Global Cartan Decomposition}
\author{Elija Perrier\thanks{University of Technology Sydney, Ultimo NSW 2007, Australia; Australian National University, Canberra. Email: \href{mailto:eper2139@uni.sydney.edu.au}{eper2139@uni.sydney.edu.au}} \and Christopher S. Jackson\thanks{Perimeter Institute, Waterloo, Ontario N2L 6B9, Canada. Email: \href{mailto:omgphysics@gmail.com}{omgphysics@gmail.com}}}

% To discuss title.

\maketitle
\begin{abstract}
Geometric methods have useful application for solving problems in a range of quantum information disciplines, including the synthesis of time-optimal unitaries in quantum control. In particular, the use of Cartan decompositions to solve problems in optimal control, especially lambda systems, has given rise to a range of techniques for solving the so-called $KP$-problem, where target unitaries belong to a semi-simple Lie group manifold $G$ whose Lie algebra admits a $\g=\k \oplus \p$ decomposition and time-optimal solutions are represented by subRiemannian geodesics synthesised via a distribution of generators in $\p$. In this paper, we propose a new method utilising global Cartan decompositions $G=KAK$ of symmetric spaces $G/K$ for generating time-optimal unitaries for targets $-iX \in [\frak{p},\frak{p}] \subset \frak{k}$ with controls $-iH(t) \in \frak{p}$. Target unitaries are parametrised as $U=kac$ where $k,c \in K$ and $a = e^{i\Theta}$ with $\Theta \in \frak{a}$. We show that the assumption of $d\Theta=0$ equates to the corresponding time-optimal unitary control problem being able to be solved analytically using variational techniques. We identify how such control problems correspond to the holonomies of a compact globally Riemannian symmetric space, where local translations are generated by $\p$ and local rotations are generated by $[\p,\p]$.\footnote{This article is a preliminary draft, put together by Elija. Chris is currently working with Elija on an updated revised draft for release in coming weeks.}
\end{abstract}

\newpage
\pagenumbering{arabic}
\tableofcontents

\section{Introduction}
Symmetry-based decompositions are a common technique for reducing problem complexity and solving constrained optimisation problems in quantum control and unitary synthesis. Among various decompositional methods, Cartan $KAK$ decompositions represent a generalised procedure for decomposing certain semi-simple Lie groups\cite{helgason_differential_1979} exhibiting involutive automorphic symmetry, akin to generalised Euler or singular-value decompositions. Cartan decompositions have found specific application across a range of domains, such as synthesising time-optimal Hamiltonians for spin-qubit systems in nuclear magnetic resonance \cite{khaneja_cartan_2001,khaneja_optimal_2005,khaneja_sub-riemannian_2002}, linear optics \cite{liu_collective_2022}, general qubit subspace decomposition \cite{earp_constructive_2005}, indirectly relating to entanglement dynamics \cite{bremner_fungible_2004} and the entangling power of unitary dynamics in multiqubit systems \cite{bullock_canonical_2004}. Other approaches in information theory \cite{drury_constructive_2008} use Cartan decompositions for quantum Shannon decompositions and quantum circuit programming \cite{tucci_introduction_2005}. More recently, their use has been proposed for efficient measurement schemes for quantum observables \cite{yen_cartan_2021}, fixed-depth Hamiltonian simulation \cite{kokcu_fixed_2021}, reducing numerical error in many-body simulations \cite{steckmann_simulating_2021} and time-reversal operators \cite{bullock_time_2005} and also measurement of quantum observables \cite{yen_cartan_2021}. 

Cartan decompositions have been of interest in quantum unitary synthesis due to the fact that multiqubit systems carrying representations of $SU(2^n)$ can often be classified as type AI and AIII symmetric spaces \cite{dalessandro_introduction_2007,khaneja_cartan_2001,noauthor_cartans_2018,su_scheme_2006,tucci_introduction_2005,khaneja_sub-riemannian_2002}. Specific interest symmetric space formalism in quantum computing has largely been due to their use in synthesising time-optimal or more efficient or controllable quantum circuits \cite{khaneja_cartan_2001,earp_constructive_2005,nielsen_geometric_2006,leifer_quantum_2008,liu_collective_2022,steckmann_simulating_2021,bullock_canonical_2004,bullock_time_2005,bullock_note_2004,gokler_efficiently_2017}. 

Symmetry-based decompositions, such as Cartan decompositions, have two-fold application in quantum control problems: firstly, symmetry-decompositions can simplify unitary synthesis via reducing the computational complexity \cite{kokcu_fixed_2022}; secondly, they have specific application in quantum control settings where the control algebra (and thus set of Hamiltonians) available are only a subalgebra of the corresponding full Lie algebra $\frak{g}$ \cite{khaneja_cartan_2001,brennen_observable_2003,swaddle_subriemannian_2017}. This paper focuses on the second use case. For unitary targets belonging to semi-simple connected groups $U_T \in G$ amenable to Cartan decomposition as $G=KAK$ (where $A < G/K$ and $K<G$), a key challenge is identifying the Hamiltonian which will synthesise the unitary optimally. While broadly utilised, Cartan-based unitary synthesis techniques have suffered from practical limitations due to exponential circuit depth \cite{nielsen_geometric_2006,huang_explicit_2007} together with difficulty in identifying the appropriate form of Cartan decomposition \cite{kokcu_fixed_2022} and form of Hamiltonian. In this paper, we address such challenges by providing a generalised procedure for time-optimal unitary and Hamiltonian synthesis using a global Cartan decomposition. Specifically, we demonstrate that for parametrised unitaries with targets in $G=KAK$, by utilising what we denote the \textit{constant}-$\theta$ method, Hamiltonians composed of generators from the horizontal (antisymmetric) bracket-generating distribution of the underlying Lie algebra associated with $G$ can, in certain cases, represent time-optimal means of synthesising such unitaries.      

\section{Symmetric spaces and KP time-optimal control}
\subsection{Overview - setting the scene}
Unitary motion is defined by the Schr\"odinger equation:
\begin{align}
\frac{dU}{dt} = -iHU \label{eq:2.1}
\end{align}
whose solution is often formally written as the time-ordered exponential:
\begin{align}
U(T) = \mathcal{T}e^{-i \int_0^T H(t)dt}U_0. \label{eq:2.2}
\end{align}
Usually, these kind of equations are considered as matrix or operator equations in a given representation. However, such equations can be considered more fundamentally differential geometric in nature. This is because time-ordered exponentials can be expanded in commutators which themselves have a structure that is often universal. Universal in this context refers to the fact that a matrix representation can be one of an infinite number of otherwise inequivalent representations. Perhaps the most useful and well-known example is angular momentum, with commutators that define the group of virtual rotations. To emphasize angular momenta and rotations as universal is to appreciate that they are concepts which precede or transcend any particular matrix representations, and this universality serves as a platform for understanding details such as energy quantization.

The differential geometry of universal structures was a famous passion of Poincaré, who originated the ideas of ``universal covering group'' and ``universal enveloping algebra'', and Klein, for whom the ``universal properties'' of category theory are an homage. In quantum physics however, although it is certainly useful to have an intuitive understanding that rotation is universal, there is usually no need to work directly with things like universal covers. Instead, the standard is to assume a Hilbert space and immediately consider angular momenta as acting on states to logically derive the spin quantum numbers and consequent matrix representations. Indeed, this quantum standard to act on states is why equation \eqref{eq:2.1} is often called ``Schrödinger.'' Followed by the basic fact that the matrix group $SU(2)$ is the fundamental representation of the universal cover of 3-dimensional rotations, one can then just use this matrix group (and perhaps a little good faith) to come to understand the universal. However, if more general transformation groups are considered, calculations with matrix representations can make the universal geometry they are representing far less apparent.

This article demonstrates that the so-called ``K-P'' problems, can be solved completely with the help of some universal techniques. The ``K-P'' problems are a family of unitary control problems where the Hamiltonian control space $ \p $ is a Lie triple system, defined by the property $[\p, [\p, \p]] \subseteq \p$, and the target Hamiltonian is in $ \k = [\p, \p] $. We expand on this below.

\subsection{KP problems for Lambda systems}
Geometric control theory is characterised by the study of the role of symmetries in optimal control problems. The $KP$-problem has been most extensively detailed by Jurdjevic in the context of geometric control theory. A common application of techniques of geometric control involves so-called \textit{lambda systems}, three-level quantum systems where only two of the levels are encoded with information of interest. In such systems, the two lowest energy states of a quantum system are coupled to the highest third energy state via electromagnetic fields \cite{dalessandro_time-optimal_2020}. The typical model for the study of lambda systems in quantum control settings is the Schr\"odinger equation in the form:
\begin{align}
    \frac{dU(t)}{dt} = \hat A U(t) + \sum_j B_j U(t) \hat u_j(t) \qquad U(0) = \mathbb{I}
\label{eqn:dal:generalschrodtimeopt}
\end{align}
where $U(t) \in SU(3)$ and $\hat A$ is a diagonal matrix comprising the energy eigenvalues of the system. The control problem becomes finding control functions $\hat u(t)$ to synthesise $U_T=U(T)$ in minimum time subject to the constraint that $||\hat u|| < C$ for constant bound $C$. The unitary targets are $U_T \in G$, a semi-simple Lie group admitting a Cartan decomposition into compact and non-compact subgroups $K$ and $P$ respectively (hence the nomenclature ``$KP$''). As discussed in \cite{albertini_symmetries_2018} and elsewhere in the literature \cite{nielsen_geometric_2006,dowling_geometry_2008}, time-optimisation can be shown to be equivalent to synthesising subRiemannian geodesics on a subRiemannian manifold implied by the homogeneous space $G/K$. In this picture, the minimum synthesis time equates to the path length of the subRiemannian curve $\gamma$ calculated by reference to the pulses $u(t)$ (see (5) in (\cite{albertini_symmetries_2018})). Intuitively, in this picture, the control Hamiltonian in $\p$ traces out a minimal-time along the manifold. For certain classes of $KP$-problem, Jurdjevic and others \cite{jurdjevic_geometric_1997,jurdjevic_hamiltonian_2001,jurdjevic_optimal_1999,dalessandro_k-p_2019} have shown that geodesic solutions to equation (\ref{eqn:dal:generalschrodtimeopt}) take a certain form (see example below).  In \cite{jurdjevic_optimal_1999} this is expressed as:
\begin{align}
    \frac{d\gamma}{dt} &=\gamma\left( e^{Kt} P e^{-At} \right)\\
    \gamma(t) &= e^{(-A + P)t}e^{-At}
    \label{eqn:jurdgeodesicsolutions}
\end{align}
with $A \in \k$ and assuming $\gamma(0)=\mathbb{I}$ for a symmetric matrix $P$ (that is in $\p$). Here $\gamma(t)$ is Discussion of the case in which $e^{(-A + P)t}$ is a scalar is set out in \cite{albertini_sub-riemannian_2020}. Detailed exposition by Jurdjevic is set out in \cite{jurdjevic_geometric_1997,jurdjevic_hamiltonian_2001,jurdjevic_optimal_1999} and elsewhere. In this paper, we show that such results can be obtained by a global Cartan decomposition in tandem with certain constraints on chosen initial conditions $U(0)=U_0$ and the choice of generators $\Phi \in \k$ used to approximate the target unitary in $K$. 
\subsection{KAK decompositions}
In many quantum control and quantum algorithm settings, the target unitary $U_T$ is an element of a semi-simple Lie group $G$ such that $U_T \in G$. In certain cases, $G$ may be decomposed into a homogeneous space $G/K$ where $K$ represents an isometry (stabilizer) subgroup $K < G$. Such semi-simple Lie groups can be decomposed as $G = KAK$ where $\a \subset \p$ and $A = \exp(\a)$, a decomposition known as a Cartan decomposition \cite{dalessandro_introduction_2007,khaneja_cartan_2000,khaneja_optimal_2005,dalessandro_lie_2008,dalessandro_k-p_2019,nielsen_optimal_2006,helgason_differential_1979}. The decomposition $G=KAK$ is considered a global Cartan decomposition (applying to the entire group) rendering $G/K$ a globally Riemannian symmetric space. The corresponding Lie algebra $\frak{g}$ may similarly be decomposed as $\frak{g} = \frak{k} \oplus \frak{p}$ where $\frak{p} = \frak{k}^\bot$. The existence of a Cartan decomposition is equivalent to the satisfaction of the following canonical Cartan commutation relations:
\begin{align}
[\frak{k},\frak{k}] \subseteq \frak{k} \qquad [\frak{p},\frak{p}] \subseteq \frak{k} \qquad [\frak{p},\frak{k}] \subseteq \frak{p}
    \label{eqn:cartancommrelations}
\end{align}
Given a choice of maximally non-compact Cartan subalgebra $\frak{a} \subset \frak{p}$, $G$ can be decomposed as $G = KAK$, where $A = e^{\frak{a}}$. Doing so allows unitaries in $G$ to be written as:
\begin{align}
    U = ke^{i\Theta}c
\end{align}
where $k,c \in K$ and $e^{i\Theta} \in A$ (where $\theta$ parametrises a generators in $\frak{a}$, e.g. a rotation angle). Satisfaction of Cartan commutation relations is also equivalent to the existence of an involution $\chi^2=\mathbb{I}$ which partitions $G$ (and $\frak{g}$) into symmetric $\chi(\frak{k})=\frak{k}$ and antisymmetric $\chi(\frak{p})=-\frak{p}$ subalgebras. Here $K = e^\frak{k},A \subset P = e^\frak{p}$ \cite{knapp_representation_2001,helgason_differential_1979,hermann_lie_1966}. Elements of $G$ can be written in terms of the relevant group of action of subgroups $K$ and $A$, including, where relevant, unitary elements of $G$. Arbitrary targets $U_T\in G$ remain reachable, however elements of the vertical  (symmetric) subalgebra must be indirectly synthesised via application of the Lie derivative (bracket) i.e. $[\frak{p},\frak{p}] \subset \frak{k}$. Such decompositions are manifestly coordinate-free when represented using differential forms. Satisfying such criteria allows $G/K$ to be equivalently characterised as a Riemannian (or subRiemannian) symmetric space. \\
\\
 We propose that for certain classes of quantum control problems, namely where the antisymmetric centralizer generators parameterised by angle $\theta$ remain constant, analytic solutions for time-optimal circuit synthesis are available for non-exceptional symmetric spaces. Such cases are explicitly where control subalgebras are limited to cases where the Hamiltonian comprises generators from a horizontal distribution (bracket-generating \cite{helgason_differential_1979,knapp_representation_2001}) $\frak{p}$ with $\frak{p}\neq \frak{g}$ (where the vertical subspace is not null). Only access to subalgebras $\frak{p} \subset \frak{g}$ is directly available. If $[\frak{p},\frak{p}] \subset \frak{k}$ holds, arbitrary generators in $\frak{k}$ may be indirectly synthesised (via application of Lie brackets) which in turn makes the entirety of $\frak{g}$ available and thus, in principle, arbitrary $U_T \in G$ (if $[\p,\p]=\k$) reachable (in a control sense).

 \subsection{Sketch of constant-$\theta$ method}
 Here we sketch the constant-$\theta$ method. In subsequent sections, we provide worked examples. A generalised form of the method is set out in Appendix (\ref{app:sec:generalmethod}). A target unitary $U_T \in G$ decomposable into a Cartan $G=KAK$ coordinate system can be written in the form:
\begin{align}
    U = k e^{i\Theta}c = q e^{-ic^\inv\Theta c}
    \label{eqn-u-kak}
\end{align}
where $c,q, k \in K$ and $e^{i\Theta} \in A \subset P$. Locally forward evolution in time can be written differentially as:
\begin{align}
    dU U^\inv = -iHdt
\end{align}
where the left-hand side consists of parameters over the manifold of unitaries (or the group $G$), while the right-hand side consists of parameters that are in a geometric sense external to the manifold, in the vector field associated with the manifold.  Taking (\ref{eqn-u-kak}) as our unitary:
\begin{align}
    dU U^\inv = k\bigg[k^\inv dk +\cos\ad_\Theta(dc c^\inv) + i\Big(d\Theta + \sin\ad_\Theta(dc c^\inv)\Big)\bigg]k^\inv
    \label{eqn:duuinvkak1}
\end{align}
Geometrically, we can interpret the left $K$ as a local frame (a choice of basis for the tangent space at each point of the group), the right $K$ as a global azimuth (a parameter that describes the position of a point on the sphere and is the same for all points along the `longitude' (or orbits of) $K$), and the $\Theta$ as the polar geodesic connecting the two (akin to latitude).
The latter two sets of parameters are the coordinates of a symmetric space. Presented in this way, the Schr\"odinger equation represents a Maurer-Cartan (differential) form \cite{helgason_differential_1979}, encoding the infinitesimal structure of the Lie group and satisfies the Maurer-Cartan equation. For our purposes, it allows us to interpret the right-hand side in terms of the relevant principal bundle connection. Intuitively, a connection provides a way to differentiate sections of the bundle and to parallel transport along curves in the base manifold $G$. The minimal connection (see (Appendix \ref{appendix:minimalconnections}) for exposition) here is a geometric way of expressing evolution only by elements in $\frak{p}$ where the quantum systems evolves according to generators in the horizontal control subalgebra of the Hamiltonian given by:
\begin{align}
    i\Big(d\Theta + \sin\ad_\Theta(dc c^\inv)\Big)
\end{align}
The evolution of the system can be framed geometrically as tracing out a path on the manifold (a Riemannian symmetric space) according to generators and amplitudes (and controls) in the Hamiltonian). This path is a sequence of quantum gates or the circuit. In quantum mechanics, we are interested in minimising the external time parameter. For this, we typically minimise the action such that the total time of a circuit evolving is:
\begin{align}
    \Omega T = \int_\gamma\!\sqrt{(idU U^\inv, idU U^\inv)} 
    %= \int\!\sqrt{(d\Theta, d\Theta) + \big(idc c^\inv, \sin^2\ad_\Theta(idc c^\inv)\big) }.
\end{align}
In common with geometric methods, the integral is usually parametrised by path length $s$ (so integrated between 0 and 1 - see the examples below) of the curve $\gamma$ traced out along the manifold. Here $\Omega = |H|$ and where $(X,Y)=\Tr(\text{ad}_X \text{ad}_Y)$
% \begin{align}
% (X,Y) = \Tr(\text{ad}_X \text{ad}_Y)/I(\lambda)
% \end{align}
is the (representation independent) Killing form which, in certain cases induces a scalar form (see below and \cite{knapp_representation_2001,cahn_semi-simple_2014}).  
As is typical in differential geometry, the minimisation of evolution time of the curve $\gamma \in G$ (where $\gamma$ represents a unitary circuit parameterised by arc length (a typical geometric path length $s$) where $s \in [0,1]$ parametrises the path from beginning to end and $ds = \Omega dt$ (see the Appendix for more discussion). To find the solution for minimal time $T$, we consider paths such that $d\Theta = 0$ (and thus $d\theta=0$) which we denote the `constant-$\theta$' method. The constant-$\theta$ method implies that for the Cartan algebra parameter $\theta$, we have $d\theta = 0$.  We conjecture that upon local variation of the total time with respect to the path, arrive at Euler-Lagrange equations which show that $dq q^\inv$ must also be constant for locally time-optimal paths.
The total time for locally time-optimal constant-$\theta$ paths becomes:
\begin{align}
    \Omega T = \min_{\theta,\phi} |\sin \text{ad}_\Theta(\Phi)|
\end{align}
where:
\begin{align}
i\Phi = \int dc c^\inv
\label{eqn:bigphiintdqqinv}
\end{align}
Determining the time-optimal geodesic path requires global variation over all geodesics, typically a hard or intractable problem. As shown in the literature \cite{nielsen_geometric_2006,dowling_geometry_2008,gu_quantum_2008,wang_quantum_2014}, variational methods can be used as a means of calculating synthesis time. For constant-$\theta$ paths, such calculations are significantly simplified as we demonstrate below.

\subsection{Holonomy targets}
The local frame has two coordinate systems that are particular, the cardinal (basis) frame ``$k$'' and the geodesic frame ``$q$.'' The cardinal frame describes the local, compact part of the group, while the geodesic frame describes the global, non-compact part of the group.
Define a target in the holonomy group $K$ to be one such that: 
\begin{align}\label{holonomy}
U(T)U(0)^\inv = e^{iX} \in K.
\end{align}
Holonomy group refers to the set of group transformations a vector undergoes when it is parallel transported around a closed loop in a manifold. For symmetric spaces, the choice of connection gives rise to an implicit subgroup $K < G$ which acts as a holonomy group. Defining the connection for the geodesic frame:
\begin{align}
q^\inv dq = c^\inv\Big[(1-\cos\ad_\Theta)(dc c^\inv)\Big]c.
\end{align}
such that the Hamiltonian does not contain elements of $\frak{k}$. The intuition for a qubit is that rotations achieved via $J_z$ are by construction parallel transporting vectors via the action of $[\p,\p]\subset \k$. Where $J_z$ is not in the control subalgebra, a way to parallel transport such vectors, achieving a $J_z$ rotation, but using other generators not in $K$ is needed. For constant-$\theta$ paths, equation (\ref{holonomy}) implies:
\begin{align}
\Ad_{\Phi} (\Theta) = \Theta
\hspace{50pt}
\text{and therefore}
\hspace{50pt}
X = (1-\cos\ad_\Theta)\Phi.
\end{align}
Note the first of these conditions is explicated as a condition that the generators comprising $\Phi \in \frak{k}$ belong to the commutant (see Appendix for discussion). Holonomic targets may be, under certain assumptions, generated via unitaries $U \in G/K$ which in a Riemannian context are paths with zero geodesic curvature, indicating parallel translation in a geometric sense. By contrast, where the manifold is subRiemannian (i.e. where $\frak{p} < \frak{g}$) then subRiemannian geodesics may exhibit non-zero geodesic curvature by comparison to $G$ and $\frak{g}$ as a whole. 

\subsection{Symmetric space controls in $\frak{p}$}
Summarising the above, our conjecture is that for symmetric space controls in $\frak{p}$ with holonomy targets in $\frak{k}$ that constant-$\theta$ paths are time-optimal. 
For such constant-$\theta$ paths, the objective becomes to calculate:
\begin{align}\label{NielsenProb}
\Omega T = \min_{\Theta,\Phi}\Big|\sin\ad_\Theta(\Phi)\Big|
\end{align}
under the constraints
\begin{align}
\Ad_{\Phi} (\Theta)= \Theta
\label{eqn:main-eiadPhiTheta=Theta}
\end{align}
and
\begin{align}
X = (1-\cos\ad_\Theta)\Phi.
\label{eqn:main-x=1-cosadthetabigphi}
\end{align}
The variations can be performed elegantly by the feature that $\Theta$ is in a symmetric space and $\Phi$ is in a reductive Lie group. From this point, the problem of minimising time is undertaken using variational techniques (such as Lagrange multipliers). The constant-$\theta$ assumption allows us to simplify this problem to a significant degree. We also show in our general exposition how a transformation to a restricted Cartan-Weyl basis can also assist in simplifying the often challenging global minimisation problem.

\subsection{Cartan decomposition}
To explicitly connect the $G=KAK$ decomposition to the typical Schr\"odinger equation, consider the $KAK$ decomposition of a unitary $U \in G$ given by:
\begin{align*}
    U = qe^a k
\end{align*}
where $q, k \in K$ and $e^a \in A$. Schrodinger's equation can be written consistent with the Maurer-Cartan form (see Appendix) as:
\begin{align}
    dU U^{-1} = -iHdt
    \label{eqn:schrodcartandecomp}
\end{align}
Expanding out we have:
\begin{align*}
    dU &= d(q e^a k) = e^a kdq + q e^a k da + qe^a dk\\
    U^{-1} &= (q e^a k)^{-1} = k^{-1}e^{-a}q^{-1}
\end{align*}
Which resolves to:
\begin{align*}
    dU U^\inv &= dq q^\inv + q da q^\inv + q e^a dk k^\inv e^{-a} q^\inv \\
    &= dqq^\inv + q da q^\inv + q e^{ad_a} (dk k^\inv) q^\inv
\end{align*}
The adjoint term can be decomposed into symmetric and anti-symmetric parts:
\begin{align*}
    e^{ad_a}(X) &= \underbrace{\cosh(ad_a)(X)}_{\text{even powers}} + \underbrace{\sinh(ad_a)(X)}_{\text{odd powers}}
\end{align*}
such that:
\begin{align}
    e^{ad_a}(dk k^{-1}) = \underbrace{\cosh(ad_a)(dk k^{-1})}_{\in \frak{k}} + \underbrace{\sinh(ad_a)(dk k^{-1})}_{\in \frak{p}} 
    \label{eqn:expcoshsinh}
\end{align}
As $\frak{a} \subset \frak{p}$ and given the Cartan commutation relations, we have $[\frak{a},\frak{k}] \subset \frak{p}$ while $[\frak{a},[\frak{a},\frak{k}]] \subset \frak{k}$. Thus the symmetric term in equation (\ref{eqn:expcoshsinh}) above, comprising even powers of generators is in $\frak{k}$ while the antisymmetric term, comprising odd powers of generators, will be in $\frak{p}$. Rearranging equation (\ref{eqn:schrodcartandecomp}):
\begin{align}
    dU U^\inv = q[da + \sinh(ad_a)(dkk^\inv) + q^\inv dq + \cosh(ad_a)(dkk^\inv)] q^\inv
    \label{eqn:dUUsinhcosh}
\end{align}
The first two terms are in $\frak{p}$ while the latter two are in $\frak{k}$. The orthogonal partitioning from the Cartan decomposition ensures simplification such that cross-terms such as Tr$(pk)$ vanish. Thus:
\begin{align}
    \text{Tr}\left[ (dUU^\inv)^2  \right] & = \text{Tr} \left[(da + \sinh(ad_a)(dkk^\inv))^2  \right] + \text{Tr}\left[ (q^\inv dq + \cosh(ad_a)(dkk^\inv))^2 \right]
\end{align}
and:
\begin{align}
    \text{Tr}\left[ (dUU^\inv)^2  \right] & = \text{Tr} \left[ da^2 \right] + \text{Tr}\left[\sinh(ad_a)(dkk^\inv))^2  \right] + \text{Tr}\left[ (q^\inv dq + \cosh(ad_a)(dkk^\inv))^2 \right]
\end{align}
using $\text{Tr}(X ad_Y(Z) = -\text{Tr}(ad_Y(X)Z)$. 

To demonstrate the constant-$\theta$ method, we assume that our controls (generators in our Hamiltonian) are in $\frak{p}$. Doing so restricts our Hamiltonian to the horizontal subspace under which the quantum state is parallel transported as the system evolves. This is equivalent to approximating a subRiemannian geodesic circuit over the relevant differentiable manifold for our chosen group $G/K$. The choice of generators that enables such parallel transport without the use of generators in $\frak{k}$ is equivalent to a Hamiltonian comprising only generators in $\frak{p}$, ensuring the quantum state undergoes parallel transport along curves $\gamma \in G/K$ such that $\nabla_{\dot{\gamma}} \dot{\gamma} = 0$.  In equation (\ref{eqn:dUUsinhcosh}) this is equivalent to setting $da + \sinh(ad_a)(dkk^\inv)=0$ or $da = -\sinh(ad_a)(dkk^\inv)$ (in geometric parlance, setting a minimal connection see Appendix (\ref{appendix:minimalconnections})).  
Using a change of variables we denote:
\begin{align}
    a = i\Theta \qquad k=e^{i\Phi} \qquad q=e^{i\Psi}
\end{align}
where $dkk^\inv = id\Phi$ and $dqq^\inv = id\Psi$. Using $\cosh(ad_{i\Theta}) = \cos(ad_\Theta)$ and $\sinh(ad_{i\Theta}) = \sin(ad_\Theta)$ the connection becomes:
\begin{align}
    d\Psi + \cos(ad_\Theta)(d\Phi) = 0
\end{align}
Assuming a constant theta $d\Theta = 0$ together with the above transformations allows our Hamiltonian in equation (\ref{eqn:dUUsinhcosh}) to be written as:
\begin{align}
    Hdt = e^{i\Psi}[-i\sin(ad_\Theta)(d\Phi)] e^{-i\Psi}
\end{align}
From this point, we must then solve the minimisation problem. 
\section{Time-optimal control examples}
We demonstrate such time-optimal analytic solutions for a few common types of symmetric space quantum control with targets in $SU(2)$ and $SU(3)$ to explicate the constant-$\theta$ method. We generalise this method in Appendix (\ref{app:sec:generalmethod}).

\subsection{$SU(2)$ time-optimal control} \label{sec:su2timeoptimalcontrol}
Consider $G=SU(2)$ with isometry group $K = S(U(1) \times U(1)) = \text{span}\{ e^{i \eta \sigma_z}\}$. Further define Pauli matrices $\{ \sigma_k \}$ with angular momenta $J_k = \frac{1}{2}\sigma_k$. Define the Cartan conjugation $\chi$ as:
\begin{align}
    U^\chi = e^{-iJ_z\pi} U^\dagger e^{iJ_z \pi}
\end{align}
together with a Cartan projection $\pi(U) = U^\chi U$. To define the relevant Cartan decomposition, we note the quotient group $G/K$ corresponds to the symmetric space:
\begin{align}
    S^2 \approx \frac{SU(2)}{S(U(1) \times U(1))} \equiv \text{AIII(1,1)}
\end{align}
where $S^2=\pi(G)=K/G$. As distinct from the projection in \cite{boozer_time-optimal_2012}, the projection is representation independent. The Cartan projection above defines the relevant $G=KAK$ decomposition. The corresponding Cartan decomposition of the Lie algebra $\frak{g}$ satisfying equations (\ref{eqn:cartancommrelations}) is:
\begin{align}
    \frak{g} = \frak{su}(2) = \underbrace{\braket{-iJ_z}}_{\frak{k}} \oplus \underbrace{\braket{-iJ_x, -iJ_y}}_{\frak{p}}
    \label{eqn:frakg=k+p}
\end{align}
with Cartan subalgebra chosen to be $\frak{a} = \braket{-iJ_y}$. We can easily see the Cartan commutation relations (equation (\ref{eqn:cartancommrelations})) hold. For $S^2$ above this corresponds to the Euler decomposition:
\begin{align}
    U = e^{iJ_z \psi} e^{iJ_y \theta} e^{iJ_z \phi}
\end{align}
Define:
\begin{align}
    k=e^{iJ_z \psi} \qquad i\Theta = iJ_y \theta \qquad c = e^{i J_z \phi} \qquad q = kc = e^{iJ_z \chi}
\label{eqn:su2:kiThetacqparams}
\end{align}
to represent the unitary as:
\begin{align}
    U = ke^{i\Theta}c = qe^{ic^\inv \Theta c}
    \label{eqn:ukeithetac}
\end{align}
Under this change of variables, the Cartan conjugation becomes:
\begin{align}
    k^\chi = k^\inv
\qquad
(e^{i\Theta})^\chi = e^{i\Theta}
\qquad
(UV)^\chi = V^\chi U^\chi
\end{align}
with (using the adjoint action):
\begin{align}
    \pi(ke^{i\Theta}c) = e^{2c^\inv i\Theta c}
\end{align}
The Schrodinger equation then becomes:
\begin{align}
    dUU^\inv
&= k\left(k^\inv dk + \cos\ad_\Theta(dcc^\inv) + id\Theta + i \sin\ad_\Theta(dcc^\inv)\right)k^\inv\\
& = k\Big(iJ_z(d\psi+d\phi\cos(\theta)) + iJ_yd\theta - iJ_xd\phi\sin(\theta)\Big)k^\inv
\label{eqn:su2mauercartan}
\end{align}
by taking the relevant differentials and inverses in equation (\ref{eqn:su2:kiThetacqparams}), using equation (\ref{eqn:econjsinhcosh}) and calculating the adjoint action on $c$. Under this Cartan decomposition, we see equation (\ref{eqn:su2mauercartan}) partitioned into symmetric $\frak{k}$ and antisymmetric $\frak{p}$ part:
\begin{align}
    dUU^\inv = k\Big(\underbrace{iJ_z(d\psi+d\phi\cos(\theta)}_{\in \frak{k}}) + \underbrace{iJ_yd\theta - iJ_xd\phi\sin(\theta)}_{\in \frak{p}}\Big)k^\inv
\end{align}
Restricting $dUU^\inv \in \frak{p}$ is equivalent to defining a minimal connection $d\psi = -d\phi \cos \theta$. That is, effectively setting the symmetric part of the Hamiltonian to zero (or in geometric parlance, restricting to the horizontal subspace corresponding to the underlying subRiemannian manifold). The Hamiltonian becomes:
\begin{align}
    H = i\frac{dU}{dt}U^\inv = k\Big(J_x\dot\phi\sin(\theta) -J_y\dot\theta \Big)k^\inv
\end{align}
where we have defined conjugate momenta $\dot\phi,\dot\theta$ for extremisation below. To calculate the optimal time, we first define the Killing form on $\frak{g}$ for which we require normalisation of the extremised action in (such that we can define an appropriately scaled norm and metric).  For a subRiemannian space of interest in the adjoint representation, the Euclidean norm is then simply defined in terms of the Killing form as $|X| = \sqrt{(X,X)}$ such that:
\begin{align}
    |idUU^\inv|^2 = (d\psi+d\phi\cos(\theta))^2 + d\theta^2 + (d\phi\sin(\theta))^2
    \label{eqn:iduuinv^2}
\end{align}
where $|J_z|^2=\mathbb{I}$. Define the energy of the Hamiltonian as $|H| = \Omega$ such that:
\begin{align}
    \Omega t = \Omega \int_\gamma dt = \int_\gamma |idUU^\inv| =  \int_\gamma \left|i\frac{dU}{ds}U^\inv\right| ds
\end{align}
Here $\gamma$ defines the curve along the manifold generated by the Hamiltonian and $t$ the time elapsed. Calculating path length here is equivalent to approximating time elapsed modulo $\Omega$. Consistent with typical differential geometric methods, we parametrise by arc length $ds$\cite{do_carmo_differential_2016}. Define:
\begin{align}
    \dot t = \left|i\frac{dU}{ds}U^\inv\right| = \frac{dt}{ds}
\end{align}
setting optimal (minimal) time as:
\begin{align}
    T = \min_\gamma t = \min \left| \int_\gamma \dot t ds  \right|
    \label{eqn:su2:t=mint}
\end{align}
Extremisation can be performed using the method of Lagrange multipliers and the minimal connection above. As we demonstrate below, doing so in conjunction with the constant-$\theta$ assumption simplifies the variational problem of estimating minimal time. The relevant action is given by:
\begin{align}
    S = \Omega t =  \int_\gamma\Big(\dot{t} + \lambda J_z(\dot\psi+\dot\phi\cos(\theta))\Big) ds
    \label{eqn:su2:s=omegataction}
\end{align}
noting again the role of the connection. Parametrising by arc length $s$ we have:
\begin{align}
    \left|i\frac{dU}{ds}U^\inv\right|^2 = (\dot\psi+\dot\phi\cos(\theta))^2 + \dot\theta^2 + (\dot\phi\sin(\theta))^2
\end{align}
Extremising the action $\delta S = 0$ resolves the canonical position and momenta via the equation above as:
\begin{align}
\Omega \frac{\delta t}{\delta \dot{\psi}} = \frac{1}{\dot{t}}\big(\dot\psi+\dot\phi\cos(\theta)\big) + \lambda
\label{eqn:lagrangedpsi}
\end{align}
\begin{align}
\Omega \frac{\delta t}{\delta \dot{\phi}} = \frac{\dot\phi}{\dot{t}} + \left(\frac{\dot\psi}{\dot{t}}+\lambda\right)\cos(\theta)
\label{eqn:lagrangedphi}
\end{align}
\begin{align}
\Omega \frac{\delta t}{\delta \dot{\theta}} = \frac{\dot\theta}{\dot{t}}
\label{eqn:lagrangedtheta}
\end{align}
\begin{align}
\Omega \frac{\delta t}{\delta \theta}
= -\left(\frac{\dot\psi}{\dot{t}}+\lambda\right)\dot\phi\sin(\theta)
\label{eqn:lagrangetheta}
\end{align}
\begin{align}
\Omega \frac{\delta t}{\delta \lambda}
=\dot\psi+\dot\phi\cos(\theta)
\label{eqn:lagrangedlambda}
\end{align}
where we have assumed vanishing quadratic infinitesimals to first order e.g. $d\phi^2=0$. We note that given $\lambda$ is constant, it does not affect the total time $T$. The choice of $\lambda$ can be considered a global gauge degree of freedom i.e. $\frac{\partial T}{\partial \lambda}=0$ (i.e. regardless of $\lambda$ minimal time $T$ remains the same). The minimal connection constraint (see Appendix (\ref{eqn:general:connection})): 
\begin{align}
k^\inv dk = - \cos\ad_\Theta(dcc^\inv) 
\label{eqn:general:connection}
\end{align}
can be written as:
\begin{align}
    \dot\psi=-\dot\phi\cos(\theta)
    \label{eqn:minimalconnpsiphi}
\end{align} 
as we have specified $k$. The connection and equation (\ref{eqn:lagrangedlambda}) imply that $\dot\psi(s)$ becomes a local gauge degree of freedom in that it can vary from point to point along the path parameter $s$ without affecting the physics of the system (i.e. the rate of change of $\psi$ can vary from point to point without affecting the energy $\Omega$ or time $T$ of the system). That is:
\begin{align}
    \frac{\delta T}{\delta \dot\psi}=0
\end{align}
We can simplify the equations of motion by setting a gauge fixing condition (sometimes called a gauge trajectory). Thus we select:
\begin{align}
    \dot \psi/\dot t + \lambda = 0
    \label{eqn:gaugefixing}
\end{align}
Recalling extremisation via $dS=0$, we find equations (\ref{eqn:lagrangedphi}) and (\ref{eqn:lagrangedtheta}) become:
\begin{align}
    dS = \Omega \frac{\delta t}{\delta \dot \phi} + \Omega\frac{\delta t}{\delta \dot \theta} = \frac{\dot \phi}{\dot t} + \frac{\dot \theta}{\dot t} = 0 
\end{align}
Thus $\dot \phi/\dot t$ and $\dot \theta/\dot t$ are constant by the constant-$\theta$ assumption i.e. $\dot\theta=0$. Minimising over constant $\dot \theta$:
\begin{align}
\Omega \frac{\partial T}{\partial \dot\theta}=(\dot\theta/\dot{t})\int_\gamma ds = 0
\end{align}
confirms the independence of $T$ from $\theta$ (i.e. as $\dot t$ and the path-length $\int_\gamma ds \neq 0$). Combining the above results reduces the integrand in equation (\ref{eqn:iduuinv^2}) to dependence on the $d\phi \sin \theta$ term (as the minimal connection condition and constant $\theta$ condition cause the first two terms to vanish). 
 Such simplifications then mean optimal time is found via minimisation over initial conditions in equation (\ref{eqn:su2:t=mint}):
\begin{align}
    T = \min_{\theta,\phi} \left| \int_\gamma d\phi \sin \theta   \right|  = \min |\phi \sin \theta| \qquad \phi = \int_\gamma d\phi
    \label{eqn:minphisintheta}
\end{align}
Note by comparison with the general form of equation (\ref{eqn:general:Tsinadthetaphi}), here the $\phi \sin \theta$ term represents $\sin \text{ad}_\Theta (dcc^\inv) = \sin \text{ad}_\Theta(\Phi)$. The above method shows how the constant-$\theta$ method simplifies the overall extremisation making the minimisation for $T$ manageable.\\
\\
Now consider holonomy targets of the form:
\begin{align}
U(T)U_0^\inv= e^{-iX} = e^{iJ_z2\eta} \in K
\end{align}
with controls only in $\frak{p} = \{iJ_x, iJ_y   \}$. By assumption $U$ is of $KAK$ form (equation (\ref{eqn:ukeithetac})):
\begin{align}
    U(T)U_0^\inv= e^{-iX} = e^{iJ_z2\eta} = ke^{i\Theta}c = qe^{ic^\inv \Theta c}
\end{align}
Choosing the initial condition as:
\begin{align}
    U_0 = e^{i\Theta}
\end{align} 
The simplest form is when $c$ resolves to identity. This in turn requires $c\in K$ to resolve to the identity:
\begin{align}
    M \equiv \Big\{c \in K : c\Theta c^\inv =\Theta \Big\} = \{\pm \mathbb{I}\}
\end{align}
Given we have a single element in $\k$,  this is equivalent $c = \exp(i J_z \phi)$ where:
\begin{align}
    \phi = 2\pi n
    \label{eqn:phi2pin}
\end{align}
for $n \in \mathbb{Z}$. In general we must optimise over choices of $n$, which in this case is simply $n=1$. From this choice of initial condition we have:
\begin{align}
    q(T)  &= k(T)c(T) = e^{-iX}\\
    q(T) &= e^{iJ_z\psi}e^{iJ_z\phi} = e^{iJ_z(\psi + \phi)} 
\end{align}
Note this is a relatively simple form of $X=(1-\cos\ad_\Theta)(\Phi)$ as $\Phi$ comprises only a single generator $J_z$. 
This condition is equivalent to:
\begin{align}
2\eta = \int_\gamma d\chi = \int_\gamma d\psi + d\phi = 2\pi n(1-\cos(\theta))
\end{align}
Here we have again used the minimal connection constraint (equation (\ref{eqn:minimalconnpsiphi})) for substitution of variables. Thus:
\begin{align}
    \cos(\theta) = 1-\frac{\eta}{n\pi}
    \label{eqn:costhetaetapi}
\end{align}
Substituting into equation (\ref{eqn:minphisintheta}) we have that:
\begin{align}
    \Omega T & =  \min_n 2\pi n\sqrt{\frac{2\eta}{n\pi}-\left(\frac{\eta}{n\pi}\right)^2}\\
& = 2\min_n \sqrt{\eta(2\pi n-\eta)}\\
& = 2 \sqrt{\eta(2\pi-\eta)}
\end{align}
where we have used:
\begin{align}
    \sin^2\theta = \left(1-\frac{\eta}{n\pi} \right)^2 = \frac{2\eta}{n\pi}-\left(\frac{\eta}{n\pi}\right)^2
\end{align}
using trigonometric identities and setting $n=1$. Note the time optimality is consistent with \cite{boozer_time-optimal_2012}, namely: 
\begin{align}
    T = \frac{2 \sqrt{\eta(2\pi-\eta)}}{\Omega}
\end{align}
We now have the optimal time in terms of the parametrised angle of rotation for $J_z$. To specify the time-optimal control Hamiltonian, recall the gauge fixing condition (equation (\ref{eqn:gaugefixing}), which can also be written $\psi/t + \lambda = 0$) such that:
\begin{align}
    \lambda & = -\dot\psi/\dot t\\
& = -\psi(T)/T\\
& = \frac{\phi}{T}\cos(\theta)\\
& = \Omega\frac{\pi-\eta}{\sqrt{\eta(2\pi-\eta)}}
\label{eqn:lambdaturningrate}
\end{align}
where we have used equations (\ref{eqn:phi2pin}) and (\ref{eqn:costhetaetapi}). Connecting the optimal time to the control pulses and Hamiltonian, note that $\lambda$ can be regarded (geometrically) as the rate of turning of the path. In particular, noting that $\lambda = -\psi(T)/T$, we can regard $\lambda$ as the infinitesimal rotation for time-step $dt$. In a control setting with a discretised Hamiltonian, we regard it as the rotation per interval $\Delta t$. Thus $-\psi(T) \to \lambda T$ and per time interval $-\psi(t) \to \lambda t$. The Hamiltonian then becomes: 
\begin{align}
H(t)= \Omega e^{-iJ_z\lambda t}e^a e^{iJ_z\lambda t} 
\end{align}
where $a \in \frak{a}\subset \frak{p}$. Selecting $J_x$, the Hamiltonian resolves to:
\begin{align}
    H(t) &= \Omega e^{-iJ_z\lambda t/2}J_x e^{iJ_z\lambda t/2} \\
    &=\frac{\Omega}{2} \begin{pmatrix}
        e^{-i \lambda t/2} & 0 \\
        0 & e^{i\lambda t/2} 
    \end{pmatrix}
    \begin{pmatrix}
        0 & 1 \\
        1 & 0
    \end{pmatrix}
    \begin{pmatrix}
        e^{i \lambda t/2} & 0 \\
        0 & e^{-i\lambda t/2} 
    \end{pmatrix}
    \\
    &= \frac{\Omega}{2} \begin{pmatrix}
        e^{-i \lambda t/2} & 0 \\
        0 & e^{i\lambda t/2} 
    \end{pmatrix}
    \begin{pmatrix}
        0 & e^{-i \lambda t/2} \\
        e^{i\lambda t/2}  & 0
    \end{pmatrix}\\
    &= \frac{\Omega}{2} \begin{pmatrix}
        0 & e^{-i \lambda t} \\
        e^{i\lambda t}  & 0
    \end{pmatrix}\\
     &= \frac{\Omega}{2} \begin{pmatrix}
        0 & \cos(\lambda t)-i \sin(\lambda t) \\
        \cos(\lambda t)+i \sin(\lambda t)  & 0
    \end{pmatrix}\\
    &= \Omega (\cos \lambda t J_x + \sin \lambda t J_y)
\end{align}
The Hamiltonian is comprised of control subalgebra generators $J_x, J_y \in \frak{p}$ with control amplitudes $\lambda$ given by their coefficients over time-interval $t$. 
% By normalising the Hamiltonian we obtain the equivalent result to \cite{boozer_time-optimal_2012}. 
In \cite{boozer_time-optimal_2012}, the target unitary is of the form:
\begin{align}
    U_T(t) = e^{i\frac{\eta}{2}\sigma_z} = e^{i\eta J_z}
\end{align}
With $\nu = 1 - \eta/(2\pi)$, the time-optimal solution for $\alpha(t)$ becomes $\alpha(t) = \omega t$ where:
\begin{align}
    \omega/\Omega = \frac{2\nu}{\sqrt{1-\nu^2}}
\end{align}
The minimum control time is given by:
\begin{align}
    T = \frac{\pi \sqrt{1-\nu^2}}{\Omega}
\end{align}

\subsection{$SU(3)/S(U(1) \times U(2))$ time-optimal control } \label{section:SU(3)}
\subsubsection{Overview}
We now consider now the constant-$\theta$ method of relevance to lambda systems, specifically for the $SU(3)/S(U(1) \times U(2))$ (AIII(3,1) type) symmetric space, distinguished by the choice of Cartan decomposition. The fundamental representation of $SU(3)$ generators is via the Gell-man matrices:
\begin{align}
\lambda_1 &= 
\begin{pmatrix} 
0 & 1 & 0 \\
1 & 0 & 0 \\
0 & 0 & 0 \\
\end{pmatrix} = \ketbra{0}{1} + \ketbra{1}{0} \quad
&
\lambda_2 &= 
\begin{pmatrix} 
0 & -i & 0 \\
i & 0 & 0 \\
0 & 0 & 0 \\
\end{pmatrix} = -i\ketbra{0}{1} + i\ketbra{1}{0} \\
\lambda_3 &= 
\begin{pmatrix} 
1 & 0 & 0 \\
0 & -1 & 0 \\
0 & 0 & 0 \\
\end{pmatrix} = \ketbra{0}{0} - \ketbra{1}{1} \quad
&
\lambda_4 &= 
\begin{pmatrix} 
0 & 0 & 1 \\
0 & 0 & 0 \\
1 & 0 & 0 \\
\end{pmatrix} = \ketbra{0}{2} + \ketbra{2}{0} \\
\lambda_5 &= 
\begin{pmatrix} 
0 & 0 & -i \\
0 & 0 & 0 \\
i & 0 & 0 \\
\end{pmatrix} = -i\ketbra{0}{2} + i\ketbra{2}{0} \quad
&
\lambda_6 &= 
\begin{pmatrix} 
0 & 0 & 0 \\
0 & 0 & 1 \\
0 & 1 & 0 \\
\end{pmatrix} = \ketbra{1}{2} + \ketbra{2}{1} \\
\lambda_7 &= 
\begin{pmatrix} 
0 & 0 & 0 \\
0 & 0 & -i \\
0 & i & 0 \\
\end{pmatrix} = -i\ketbra{1}{2} + i\ketbra{2}{1} \quad
&
\lambda_8 &= 
\frac{1}{\sqrt{3}}
\begin{pmatrix} 
1 & 0 & 0 \\
0 & 1 & 0 \\
0 & 0 & -2 \\
\end{pmatrix} = \frac{1}{\sqrt{3}}(\ketbra{0}{0} + \ketbra{1}{1} - 2\ketbra{2}{2})
\end{align}
Following \cite{cahn_semi-simple_2014,byrd_geometry_1997}, we set out commutation relations for the adjoint representation of $\frak{su}(3)$, the Lie algebra of $SU(3)$ in Table (\ref{tab:su3commutationHIII}). The row label indicates the first entry in the commutator, the column indicates the second. 
%======= Commutation Table (Main SU3)
%=========COMMUTATION TABLE (BASE)
\begin{center}
    \begin{table}[h!]
    \centering
    \fontsize{8pt}{8pt}
    \begin{tabular}{ |c|c|c|c|c|c|c|c|c|  } 
 \hline
  & ${\color{red}-i\lambda_1}$ & ${\color{red}-i\lambda_2}$ & ${\color{red}-iH_{\text{III}}}$ & ${\color{red}-iH_{\text{III}}^\perp}$ & $-i\lambda_4 $& $-i\lambda_5 $ & $-i\lambda_6$ & $-i\lambda_7$ \\ 
 \hline
%\lambda1
 ${\color{red}-i\lambda_1}$ & \cellcolor{matteYellow}0 &\cellcolor{matteYellow} ${\color{red}i\sqrt{3}H_{\text{III}} -iH_{\text{III}}^\perp}$  & \cellcolor{matteYellow}${\color{red}-i\sqrt{3}\lambda_2}$ &\cellcolor{matteYellow} ${\color{red}i\lambda_2}$ &  $-i\lambda_7$ &  $i\lambda_6$ & $-i\lambda_5$ & $i\lambda_4$ \\
 \hline

%\lambda2
${\color{red}-i\lambda_2}$ &\cellcolor{matteYellow} ${\color{red}-i\sqrt{3}H_{\text{III}} +iH_{\text{III}}^\perp}$ &\cellcolor{matteYellow} 0  & \cellcolor{matteYellow}${\color{red}\sqrt{3}i\lambda_1}$ &  \cellcolor{matteYellow}${\color{red}-i\lambda_1}$ &  $-i\lambda_6$ &  $-i\lambda_7$ & $i\lambda_4$ & $i\lambda_5$ \\ 
\hline

%===H_{\text{III}}
 ${\color{red}-iH_{\text{III}}}$ &\cellcolor{matteYellow} ${\color{red}i\sqrt{3}\lambda_2}$ &\cellcolor{matteYellow} ${\color{red}-i\sqrt{3}\lambda_1}$ &\cellcolor{matteYellow} $0$ &\cellcolor{matteYellow} $0$ & $0$ & $0$ &  $-i\sqrt{3}\lambda_7$ &  $i\sqrt{3}\lambda_6$ \\ 
 \hline

 % ===H_{\text{III}}^\perp
${\color{red}-iH_{\text{III}}^\perp}$ &\cellcolor{matteYellow} ${\color{red}-i\lambda_2}$ &\cellcolor{matteYellow} ${\color{red}i\lambda_1}$ &\cellcolor{matteYellow} $0$ &\cellcolor{matteYellow} $0$& $-i2\lambda_5$ & $i2\lambda_4$ &  $-i\lambda_7$ & $i\lambda_6$ \\ 
 \hline

 %\lambda4
 $-i\lambda_4 $ &  $i\lambda_7$ &  $i\lambda_6$ &$0$ & $i2\lambda_5$ &\cellcolor{matteGreen}  0 & \cellcolor{matteGreen}${\color{red}-2iH_{\text{III}}^\perp}$ &\cellcolor{matteGreen} ${\color{red}-i\lambda_2}$ &\cellcolor{matteGreen} ${\color{red}-i\lambda_1}$ \\
 \hline

 %\lambda5
$-i\lambda_5 $ &  $-i\lambda_6$ &  $i\lambda_7$ &$0$&$-i2\lambda_4$ & 
 \cellcolor{matteGreen} ${\color{red}i2H_{\text{III}}^\perp}$ &\cellcolor{matteGreen} $0$ &\cellcolor{matteGreen} ${\color{red}i\lambda_1}$ &\cellcolor{matteGreen} ${\color{red}i\lambda_2}$ \\ 
 \hline

 %\lambda6
 $-i\lambda_6$ & $i\lambda_5$ & $-i\lambda_4$ &$i\sqrt{3}\lambda_7$&$i\lambda_7$ &\cellcolor{matteGreen} ${\color{red}i\lambda_2}$ &\cellcolor{matteGreen} ${\color{red}-i\lambda_1}$ &\cellcolor{matteGreen} 0 &\cellcolor{matteGreen} ${\color{red}-i\left(\sqrt{3}H_{\text{III}} + H_{\text{III}}^\perp\right)}$ \\ 
 \hline

 %\lambda7
 $-i\lambda_7$ & $-i\lambda_4$ & $-i\lambda_5$ &$-i\sqrt{3}\lambda_6$&$-i\lambda_6$ &\cellcolor{matteGreen} ${\color{red}i\lambda_1}$ &\cellcolor{matteGreen} ${\color{red}-i\lambda_2}$ &\cellcolor{matteGreen} ${\color{red}i\left(\sqrt{3}H_{\text{III}} + H_{\text{III}}^\perp\right)}$ &\cellcolor{matteGreen} 0 \\
 \hline 
\end{tabular}
    \caption{Commutation relations for generators in adjoint representation of $\frak{su}(3)$. The Cartan decomposition is $\frak{g} = \frak{k} \oplus \frak{p}$ where $\frak{k} = \spana\{-i\lambda_1,-i\lambda_2,-iH_{\text{III}},-iH_{\text{III}}^\perp\}$ (red) and $\frak{p} = \spana\{-i\lambda_4, -i\lambda_5, -i\lambda_6, -i\lambda_7\}$ (black). As can be seen visually, the decomposition satisfies the Cartan commutation relations (equation (\ref{eqn:cartancommrelations})): the yellow region indicates $[\frak{k},\frak\k] \subset \frak{k}$, the green region that $[\frak{p},\frak{p}] \subset \frak{k}$ and the white region that $[\frak{p},\frak{k}] \subset \frak{p}$. From a control perspective, by inspection is clear that any element in $\frak{k}$ can be synthesised (is reachable) via linear compositions of the adjoint action of $\frak{p}$ upon itself (the green region). We choose $\frak{a} = \braket{-i\lambda_5}$ with $\h = \braket{-iH_{\text{III}},-i\lambda_5}$.}
    \label{tab:su3commutationHIII}
\end{table}
\end{center}
%=============end table
The typical Cartan decomposition of $\frak{g}=\frak{su}(3)=\k \oplus \p$ follows: 
\begin{align}
    \frak{k}&=\text{span}\{-i\lambda_1,-i\lambda_2,-i\lambda_3,-i\sqrt{3}\lambda_8\}\\
    \frak{p}&=\text{span}\{-i\lambda_4, -i\lambda_5, -i\lambda_6, -i\lambda_7\}
\end{align}
with a maximally compact Cartan subalgebra for $\frak{g}$ often chosen to be $\frak{h}=\text{span}\{-i\lambda_3,-i\sqrt{3}\lambda_8\}$ At this stage, our Cartan subalgebra is not maximally non-compact (having no intersection with $\p$), which we deal with below. Our interest is in $\Lambda$ systems, that is, 3-level quantum systems where the Hamiltonian control space is a 4-dimensional space of optical transitions generated by $\p$ (where $(\lambda_1,\lambda2)$ may be substituted for $(\lambda_6,\lambda_7)$. The target space is the 4-dimensional unitary subgroup $K = S(U(1) \times U(2))$ of microwave transitions generated by $\k$. With respect to this decomposition, it is convenient to introduce a slight change of basis:
\begin{align}
    H_{\text{III}} &= -\frac{\sqrt{3}}{2}\lambda_3 +\frac{1}{2}\lambda_8 = \frac{1}{\sqrt{3}}(-\ketbra{0}{0} + 2\ketbra{1}{1} - \ketbra{2}{2}) 
\\ H_{\text{III}}^\perp &= \frac{1}{2}\lambda_3 + \frac{\sqrt{3}}{2}\lambda_8 = \ketbra{0}{0} - \ketbra{2}{2}
\end{align}
Note for convenience:
\begin{align}
\lambda_3 &=  -\frac{\sqrt{3}}{2}H_{\text{III}} + \frac{1}{2}H_{\text{III}}^\perp \qquad \lambda_8 = \frac{\sqrt{3}}{2}H_{\text{III}}^\perp + \frac{1}{2}H_{\text{III}}\\
\lambda_3 + \sqrt{3}\lambda_8 &= 2H_{\text{III}}^\perp \qquad -\lambda_3 + \sqrt{3}\lambda_8 = \sqrt{3}H_{\text{III}} + H_{\text{III}}^\perp
\end{align}
and:
\begin{align}
    \ad_\Theta^2 (-iH_{\text{III}}) &= 0 \qquad \ad_\Theta^2 (-iH_{\text{III}}^\perp) = (-2\theta)^{2}(-H_{\text{III}}^\perp)
\end{align}
Under this change of basis:
\begin{align}
    \frak{k}&=\text{span}\{-i\lambda_1,-i\lambda_2,-iH_{\text{III}},-iH_{\text{III}}^\perp\}\\
    \frak{p}&=\text{span}\{-i\lambda_4, -i\lambda_5, -i\lambda_6, -i\lambda_7\}
    \label{eqn:su3:kandphIIIetc}
\end{align}
Under this transformation, the maximally compact Cartan subalgebra $\h=\{H_{\text{III}},H_{\text{III}}^\perp\} \in k$ in (\ref{eqn:su3:kandphIIIetc}) is entirely within $\k$. In principle, to obtain a maximally noncompact Cartan subalgebra, we conjugate $\h$ via an appropriately chosen group element (see Section 7 of Part VI of \cite{knapp_lie_1996}). However, in our case we can simply read off the combination $\{-iH_{\text{III}},-i\lambda_5\}$ (such that $\h \cap \p = -i\lambda_5$) from the commutation table above.
%=============optimal dicussion
The commutant of $\a = \spana\{-i\lambda_5\}$ is the subgroup:
\begin{align}
    M \equiv \big\{ k \in K : \forall i\Theta \in \a, k \Theta k^\inv \in \Theta \big\} = e^{\m} 
\end{align}
where in this case:
\begin{align}
    \m = \spana\{ iH_{\text{III}}\}
\end{align}
We also define:
\begin{align}
    H_{\text{III}}^{\perp''} = - \frac{\sqrt{3}H_{\text{III}} + H_{\text{III}}^\perp}{2} = \begin{pmatrix}
        0 & 0 & 0\\
        0 & 1 & 0\\
        0 & 0 & -1
    \end{pmatrix} 
\end{align}
which together with $\lambda_6$ and $\lambda_7$ define the microwave Pauli operators and the generator:
\begin{align}
    H_{\text{III}}'' = \frac{-3H_{\text{III}}^\perp - \sqrt{3 }H_{\text{III}}}{2} = \begin{pmatrix}
        -2 & 0 & 0\\
        0 & 1 & 0\\
        0 & 0 & 1
    \end{pmatrix} 
\end{align}
which commutes with the microwave qubit. Continuing, via equations (\ref{eqn:general:optimaltime}-\ref{eqn:general:optimaltimeconstraints}) we have optimal time in the form:
\begin{align}
    \Omega T = \min_{\Theta,\Phi}|\sin \ad_\Theta (\Phi)| 
    \label{eqn:su3:optimaltimesine}
\end{align}
with initial condition $U_0$=$e^{-i\Theta} \in A$ and target unitary is given by $U_T = e^{-iX}$:
\begin{align}
     X = (1-\cos \ad_\Theta)(\Phi)
\end{align}
Here $\Phi \in \k$ and is in the commutant with respect to $\Theta$. The time-optimal circuit is generated by the Hamiltonian:
\begin{align}
H(t)= e^{-i\Lambda t}\sin\ad_{\Theta_*}(\Phi_*)e^{i\Lambda t} \qquad \Lambda = \frac{\cos \ad_{\Theta_*}(\Phi_*)}{T} \equiv \cos \ad_{\dot \Theta_*}(\Delta \Phi_*)
\end{align}
Here $\Theta_*,\Phi_*$ reflects a choice of parameters (e.g. $\theta^k,\phi^j$) which minimise time $T$. 
To calculate the optimal time and corresponding Hamiltonian explicitly, we note  $\ad_\Theta (\Phi) = [\Theta,\Phi] \in \frak{p}$ while $\ad_\Theta^2(\Phi) = [\Theta,[\Theta,\Phi]]  \in \frak{k}$ (see Appendix (\ref{app:su3commutrelat})). 

\subsubsection*{General form of target unitaries in SU(3)}
We set out the general form of targets using the particular Cartan decomposition and choice of basis for SU(3) above. We set:
\begin{align}
    \Phi &= -i(\phi_1 \lambda_1 + \phi_2 \lambda_2 + \phi_3 H_{\text{III}}^{\perp''}   + \phi_4 \sqrt{3}H_{\text{III}})
\end{align}
The $\cos(\ad_\Theta)$ term is proportional to the application of $\ad_\Theta^{2}$:
\begin{align}
    \cos \ad_\Theta (\Phi) &=\cos\alpha(\theta)(\Phi)\\
    &=\cos(\theta)(-i\lambda_1) + \cos(\theta)(-i\lambda_2) + \cos(2\theta)(-iH_{\text{III}}^{\perp''}) -i \sqrt{3}H_{\text{III}}
\label{eqn:su3:cosalphaexpansion}
\end{align}
using $\cos(-\theta)=\cos(\theta)$, that $\cos\alpha(\theta)=\cos(0)=1$ for $-iH_{\text{III}}$.  Our targets are in this case of the form:
\begin{align}
    X &= (1-\cos \ad_{\Theta}) (\Phi) = (1-\cos \alpha(\theta)) (\Phi)\\
    &=-i\bigg((1-\cos(\theta))\phi_1 \lambda_1 + (1-\cos(\theta))\phi_2 \lambda_2 + (1-\cos(2\theta)) \phi_3 H_{\text{III}}^{\perp''} \bigg)
    \label{eqn:SU3:X=i(phik-costheta)Hk}
\end{align}
where $1-\cos\alpha(\theta)=0$ for $H_{\text{III}}$. Note that for more general targets involving linear combinations of $H_{\text{III}}$, specific choices or transformations of $\Phi$ and transformations of $\Theta$ such that the $H_{\text{III}}$ term does not vanish may be required (we save such discussion for later work). As per the general method set out in the Appendix, this form of $X$ can be derived as follows. We begin with the most general form of target:
\begin{align}
    X = -i(\eta_1 \lambda_1 + \eta_2 \lambda_2 + \eta_3 H_{\text{III}}^{\perp''} + \eta_4 \sqrt{3}H_{\text{III}})
\end{align}
Minimising evolution time firstly requires a choice of an initial condition. Our target can be written:
\begin{align}
    U(T)U_0^\inv=e^{-iX}=qe^{ic^\inv\Theta c}
\end{align}
with $U_0=e^{i\Theta}$ and the commutant condition we have that:
\begin{align}
    q(T) = k(T)c(T)=e^{\Phi'}e^{\Phi} =e^{\Phi''}= e^{-iX}
\end{align}
In general, we must choose $\Phi$ such that it is in the commutant:
\begin{align}
    \Phi = -i(\phi_1 \lambda_1 + \phi_2 \lambda_2 + \phi_3 \sqrt{3}H_{\text{III}} + \phi_4 H_{\text{III}}^{\perp''}) \in M
\end{align}
Where a target does not comprise a generators, its coefficient may be set to zero. Gathering terms:
\begin{align}
    q(T) &= e^{-i\Phi''}\\
    \Phi'&=-i(\psi_1 \lambda_1 + \psi_2 \lambda_2 + \psi_3 H_{\text{III}}^\perp + \psi_4 \sqrt{3} H_{\text{III}})\\
    \Phi''&=-i\bigg((\psi_1 +\phi_1) \lambda_1 + (\psi_2+\phi_2) \lambda_2 \\
    &+ (\psi_3+\phi_3) H_{\text{III}}^{\perp''} + (\psi_4 + \phi_4) \sqrt{3}H_{\text{III}} \bigg) 
\end{align}
 Using the form of minimal connection (equation (\ref{eqn:general:connection})):
\begin{align}
    \dot\psi_k = -\dot\phi_k  \cos\alpha(\theta)
\end{align}
equate coefficients for our target Hamiltonian $X$, where, by gathering terms by generator, we note that $\eta_k = \phi_k(1-(\cos\alpha_k(\theta))$ (excluding the vanishing $-iH_{\text{III}}$ term):
% \begin{align}
    % \eta_k &= \int_\gamma d\psi_k + d\phi_k = \int_\gamma d\phi_k(1-\cos\alpha_k(\theta)) = 2\pi n_k(1-(\cos\alpha_k(\theta))
    % \eta_2 &= \int_\gamma d\psi_3 + d\phi_3 = \int_\gamma d\phi_3(1-\cos(\theta)) = 2\pi n_3(1-\cos(\theta))
% \end{align}
% explicitly:
\begin{align}
    \eta_1 &=\phi_1(1-\cos(\theta)) \qquad \eta_2 =\phi_2(1-\cos(\theta))\\
    \eta_3 &=\phi_3 (1-\cos(2\theta)) \qquad \eta_4 =\phi_4 (1-1)=0
\end{align}
recovering the form in equation (\ref{eqn:SU3:X=i(phik-costheta)Hk}) up to relevant generators. Continuing:
\begin{align}
   \cos(\theta)  &= 1-\frac{1}{2}\left(\frac{\eta_1}{\phi_1} +\frac{\eta_2}{\phi_2} \right) \qquad \sin(\theta) = \sqrt{1-\cos(\theta)^2}\\
   \cos(2\theta)  &= 1-\frac{\eta_3}{\phi_3} \qquad \sin(2\theta) = \sqrt{1-\cos(2\theta)^2}
   \label{eqn:su3:costhetacos2theta}
\end{align}
Note that $\eta_4$ does not contribute. Optimal time is parametrised as:
\begin{align}
    \Omega T &= \min_{\Theta,\Phi}|\sin \ad_\Theta (\Phi)| = \min_{\theta,\phi_k} |\sin \ad_{\Theta}(\Phi)|\\
    &=\min_{\theta,\phi_k} \sqrt{\left(\sum_k \phi^2_k \sin^2\alpha_k(\theta)\right)}\\
    &= \min_{\phi_k}\sqrt{ \phi_1^2\sin^2(\theta) + \phi_2^2\sin^2(\theta)+ \phi_3^2\sin^2(2\theta) }
\end{align}
where the minimisation depends on the choice of $\phi_k$. The choice of $\phi_i$ must be such that the commutant condition $e^{i\Phi} \in M$ is satisfied. For certain targets $X$, this means:
\begin{align}
    \Phi =-i \phi_3 \sqrt{3}H_{\text{III}}  -i2\pi k(n^x \lambda_1 + n^y \lambda_2 + n^z H_{\text{III}}^{\perp''}) \in M
\end{align}
for any $\phi_3 \in \Real, k \in \mathbb{Z}$ and unit vector $\hat n = (n^x, n^y, n^z)$. The remainders of $\p$ and $\k$ pair into root spaces of $\a$:
\begin{align*}
    [i\lambda_5, i\lambda_1] &= -i\lambda_6 & [i\lambda_5, i\lambda_6] &= i\lambda_1, \\
    [i\lambda_5, i\lambda_2] &= i\lambda_7  & [i\lambda_5, i\lambda_7] &= -i\lambda_2,
\end{align*}
and
\begin{align*}
    [i\lambda_5, i\lambda_4] &= 2i\lambda_3 & [i\lambda_5, i\lambda_3] &= -2i\lambda_4.
\end{align*}
In this formulation, for a target unitary $U(T) = e^{iX}$:
\begin{align}
    X &= (1-\cos \ad_\Theta)(\Phi)\\
    &=-i2\pi k((1-\cos\theta)(n^x \lambda_1 + n^y \lambda_2) + (1-\cos(2\theta))n^z H_{\text{III}}^{\perp''}
\end{align}
For simple targets, such as those dealt with in the next section, $\phi_k$ simply be an integer multiple $2\pi k$, in which case the minimisation problem becomes one of selecting the appropriate choice of $k$ that minimises $T$. Note that as discussed, this particular form of Hamiltonian cannot reach targets in $H_{\text{III}}$. Having determined $T$, the optimal time Hamiltonian is constructed as follows. Recall from equation (\ref{eqn:general:lambda=cosadthetaphi/T}) that:
\begin{align}
H(t)&= e^{-i\Lambda t}\sin\ad_{\Theta_*}(\Phi_*)e^{i\Lambda t}\\
\Lambda & = \frac{\cos\ad_{\Theta_*}(\Phi_*)}{T}
\end{align}
noting for completeness that $\Lambda \in \frak{k}$ (and $\sin \ad_\Theta(\Phi) \in \p$). We demonstrate the technique for a specific example gate in the literature below.
\subsubsection{Comparison with existing methods}
In this section, we apply the constant $\theta$ method to derive time optimal results from \cite{albertini_sub-riemannian_2020}. The $KP$ decomposition for D'Alessandro et al. in \cite{dalessandro_time-optimal_2020} and \cite{albertini_sub-riemannian_2020} is:
\begin{align}
    \frak{p}&=\frac{1}{\sqrt{2}}\text{span}\{i\lambda_1,i\lambda_2,i\lambda_4, i\lambda_5\}\\
    \frak{k}&=\frac{1}{\sqrt{2}}\text{span}\{i\lambda_3, i\lambda_6, i\lambda_7, i\lambda_8\}
    \label{examples:dalkp}
\end{align}
As the only difference with our chosen $KP$ decomposition (up to the constant $-1/\sqrt{2}$) is swapping $-i\lambda_1,-i\lambda_2 \in \p$ and $-i\lambda_6,-i\lambda_7 \in \k$ we can use the same change of basis from $-i\lambda_3, -i\lambda_8$ to $-iH_{\text{III}},-iH_{\text{III}}^\perp$ and choice of $\a = \spana\{-i\lambda_5\}$. For convenience and continuity with \cite{albertini_sub-riemannian_2020}, we use the standard notation from that paper. The key point for our method is the choice of a Cartan subalgebra that is maximally non-compact allowing us to select $\Theta$ proportional to $-i\lambda_5$.  The form of targets in \cite{albertini_sub-riemannian_2020} $U_f$ is:
\begin{align}
    U_f &= \begin{pmatrix}
        e^{-i\phi_s} & 0 \\
        0 & \hat U_s
    \end{pmatrix}
\end{align}
where $U_s \in U(2)$ and $\det U_s = e^{i\phi_s}$. For the given $KP$ decomposition, matrices $K$ (block diagonal) and $P$ (block anti-diagonal) have the form:
\begin{align}
    K = \begin{pmatrix}
        if & 0\\
        0 & Q
    \end{pmatrix}
    \qquad 
    P = \begin{pmatrix}
        0 & \alpha & \beta \\
        -\alpha^* & 0 & 0\\
        -\beta^* & 0 & 0
    \end{pmatrix}
\end{align} 
chosen in order to eliminate the drift term. The general form of $A \in \k$ and $P =i\lambda_1 \in \p$ in that work are:
\begin{align}
    A&= \begin{pmatrix}
        a + b & 0 & 0\\
        0 & -a & -c \\
        0 &  -c & -b
    \end{pmatrix} \qquad P = i\begin{pmatrix}
        0 & 1 & 0\\
        1 & 0 & 0\\
        0 & 0 & 0
    \end{pmatrix}
\end{align}
$P$ is expressed to be an element of a Cartan subalgebra of $\p$. However the full Cartan subalgebra is not given. In that paper, the given target is a simple Hadamard gate:
\begin{align}
    U_f &=\begin{pmatrix}
        1 & 0 & \\
        0 & \frac{1}{\sqrt{2}} & \frac{i}{\sqrt{2}}\\
        0 & \frac{i}{\sqrt{2}} & \frac{1}{\sqrt{2}}
    \end{pmatrix} = \exp(H) = \exp\left(-\frac{i \pi}{4} \lambda_6\right)
\end{align}
resulting in:
\begin{align}
A = \begin{pmatrix}
    0 & 0 & 0 \\
 0 & 0 & \pm\frac{7 i}{\sqrt{15}} \\
 0 & \pm\frac{7 i}{\sqrt{15}} & 0
\end{pmatrix} \qquad P = \begin{pmatrix}
    0 & i & 0 \\
 i & 0 & 0 \\
 0 & 0 & 0 
\end{pmatrix}
\label{eqn:dalalbertiniAP}
\end{align}
In \cite{albertini_sub-riemannian_2020}, the solution to subRiemannian geodesics  (drawing upon results originally presented by Jurdjevic in \cite{jurdjevic_optimal_1999} (p.257) and later with more exposition in \cite{jurdjevic_hamiltonian_2001} (p.28)) relying upon, as Jurdjevic notes, the right invariance of the vector field under the action of elements of $K$) is of the form:
\begin{align}
    U_f &= \exp(A t) \exp((-A + P)t) = \exp(A t)
    \label{eqn:daljurdgeodesicsol}
\end{align}
In \cite{albertini_sub-riemannian_2020}, the results from equations (\ref{eqn:jurdgeodesicsolutions}) are assumed and expressed via the constraint $\exp((-A + P)t) = \exp(2\pi k/3) = \mathbb{I}$ [(where $k \in \mathbb{Z}$] and $t = t_{\min} = 2\pi T = \sqrt{15}\pi/4$, in which case:
\begin{align}
    \exp(A \sqrt{15}\pi/4) 
    &=\exp\left(\frac{\sqrt{15}\pi}{4}\begin{pmatrix}
    0 & 0 & 0 \\
 0 & 0 & \pm\frac{7 i}{\sqrt{15}} \\
 0 & \pm\frac{7 i}{\sqrt{15}} & 0
\end{pmatrix}\right) \label{eqn:dalbertini:expAT}\\
& = \begin{pmatrix}
        1 & 0\\
        0 & \exp(- 7\pi/4 \sigma_x)  
    \end{pmatrix}
    \\
    &= \begin{pmatrix}
            1 & 0 & 0\\
            0 & \cos(-7\pi/4) & i\sin(-7\pi/4)\\
            0 & i\sin(-7\pi/4) & \cos(-7\pi/4)\\
        \end{pmatrix} = U_f
\end{align}
where we choose $c = -\frac{7 i}{\sqrt{15}}$ in the second line from equation (\ref{eqn:dalalbertiniAP}) above. If the positive value is used, a similarity transformation $K \in \exp(\k)$ of the $KAK$ decomposition $\exp(A t)P \exp(-A t)$ is required. In the general form of solution to subRiemannian geodesic equations from Jurdjevic et al., the control algebra is related to the Hamiltonian via:
\begin{align}
    \exp(A t)P \exp(-A t) = \sum_j u(t)_j B_j
    \label{eqn:dalbertini:expAtPexp-At}
\end{align}
for $B_j \in \p$ and $u_j(t) \in \Real$ (with $||u|| =M $ for $M = \Omega$ constant, by the Pontryagin `bang bang' principle).
%
% , using equation (\ref{eqn:dalalbertiniAP}):
% \begin{align}
%    \exp(A t)P \exp(-A t) &= \sum_j u(t)_j B_j \\
%    &=  \frac{i}{\sqrt{2}}(u_1(t)\lambda_1 + u_5(t)\lambda_5) = \begin{pmatrix}
%          0 & \frac{i}{\sqrt{2}} u_1(t)  & \frac{1}{\sqrt{2}} u_2(t)  \\
% \frac{i}{\sqrt{2}} u_1(t)  & 0 & 0 \\
%  -\frac{1}{\sqrt{2}} u_2(t)  & 0 & 0 \\
%     \end{pmatrix}
% \label{eqn:dalbertinieHam} \\
% &= \frac{\sqrt{15}\pi}{4}\begin{pmatrix}
%         0 & \frac{i}{\sqrt{2}} & \frac{1}{\sqrt{2}} \\
%  \frac{i}{\sqrt{2}} & 0 & 0 \\
%  -\frac{1}{\sqrt{2}} & 0 & 0 \\
%     \end{pmatrix}
% \end{align}
% This is not the final form of Hamiltonian in terms of $\p$ required to reach the target, however. As we demonstrate below, a further transformation by an element in $\p$ is required to obtain $A$. 
%
\subsubsection{Constant-$\theta$ method}
To demonstrate our constant $\theta$ method, we first obtain our target generators in terms of $\k$:
\begin{align}
    \log(U_f) = \begin{pmatrix}
        0 & 0 & 0 \\
 0 & 0 & \frac{i \pi }{4} \\
 0 & \frac{i \pi }{4} & 0 \\
    \end{pmatrix} = \frac{i\pi}{4}\lambda_6 = \left( \frac{\pi}{2\sqrt{2}} \right) \left(-\frac{i}{\sqrt{2}} \right)\lambda_6 
\end{align}
which becomes our target $X$:
\begin{align}
    X &= \left( \frac{\pi}{2\sqrt{2}} \right) \left(-\frac{i}{\sqrt{2}} \right)\lambda_6 = \frac{\pi}{4}(-i \lambda_6) = -i \eta_6 \lambda_6\\
    \Phi &=-i\phi_6 \lambda_6
\end{align}
In this relatively simple example, we can choose $\Phi$ to solely consist of $-i\lambda_6$. Setting $\Theta = -i\lambda_5$ and noting $\cos\ad_\Theta(\Phi) = \cos\alpha(\theta)(\Phi)$:
\begin{align}
    X &= (1-\cos\ad_\Theta)(\Phi)\\
    &=(1-\cos\alpha(\theta))(\Phi)\\
    &=(1-\cos(\theta))(-i\phi_6\lambda_6)\\
    &= -i\eta_6 \lambda_6
\end{align}
where we have used:
\begin{align}
    \cos \ad_\Theta(\Phi) &=  \cos \alpha(\theta) \ad^{2k}_{-i\lambda_5}(-i\lambda_6) = i\cos(\theta)\lambda_6 \qquad \cos\alpha(\theta) = \cos(\theta)\\
    \sin \ad_\Theta(\Phi) & =  \sin \alpha(\theta) \ad^{2k+1}_{-i\lambda_5}(-i\lambda_6) = i\sin(\theta)\lambda_1 \qquad \sin\alpha(\theta) = \sin(\theta)
    \label{eqn:dalcosadsinad}
\end{align}
Noting $\Ad(\exp(X))(Y)=\exp(\ad_X(Y))$ such that for $c=\exp(-i\phi_6 \lambda_6)$:
\begin{align}
    e^{-i\phi_6 \lambda_6} (-i\lambda_5) e^{i\phi_6 \lambda_6} = -i\lambda_5
    % = e^{[-i\phi_6 \lambda_6,-i\lambda_5]} = \mathbb{I} 
\end{align}
our commutant condition is satisfied for $\phi_6 = 2\pi$, then:
\begin{align}
    \frac{\pi}{4} &= 2\pi(1-\cos(\theta))\\
    \cos(\theta) &= 1 - \frac{1}{8} = \frac{7}{8}\\
    \sin(\theta) &= \sqrt{1 - \left(\frac{7}{8} \right)^2} = \frac{\sqrt{15}}{8}
\end{align}
which is ``minimum T'' in \cite{albertini_sub-riemannian_2020}. Minimal time is then:
\begin{align}
    \Omega T &= \min_{\Theta,\Phi} \big|\sin\ad_\Theta(\Phi)\big|\\
    &=\min_{\theta,\phi_k} \sqrt{\phi^2_6 \sin^2\alpha(\theta)}\\
    &=\pm 2\pi \sin(\theta)\\
    &=\pm 2\pi \frac{\sqrt{15}}{8} = \pm \frac{\sqrt{15}\pi}{4} \label{eqn:dalbertiniOmegaT}
\end{align}
which is (as time must be positive) the minimum time $t$ to reach the equivalence class of target Hamiltonians (that generate $U_f$ up to conjugation) in \cite{albertini_sub-riemannian_2020}. Note in this paper we denote this minimum time to reach the target as $T$. For convenience we assume $\Omega = 1$ such that $T=\sqrt{15}\pi/4$. To calculate the Hamiltonian:
\begin{align}
H(t)= e^{-i\Lambda t}\sin\ad_{\Theta_*}(\Phi_*)e^{i\Lambda t}
\end{align}
and apply our formulation:
\begin{align}
\Lambda & = \frac{\cos\ad_{\Theta_*}(\Phi_*)}{T}\\
&=\frac{2\pi\cos\theta}{T}(-i\lambda_6)\\
&= \frac{7/8}{\sqrt{15}/8}(-i\lambda_6)\\
&= \frac{7}{\sqrt{15}}(-i\lambda_6)\\
&=\begin{pmatrix}
    0 & 0 & 0\\
    0 & 0 & -\frac{7i}{\sqrt{15}}\\
    0 & -\frac{7i}{\sqrt{15}} & 0
\end{pmatrix}
\end{align}
which is $A$ in \cite{albertini_sub-riemannian_2020}. Then from equations (\ref{eqn:dalcosadsinad}) and (\ref{eqn:dalbertiniOmegaT}):
\begin{align}
    \sin\ad_\Theta(\Phi) = i2\pi\sin(\theta)\lambda_1 &=2\pi \sin(\theta)\begin{pmatrix}
        0 & i & 0 \\
        i & 0 & 0\\
        0 & 0 & 0
    \end{pmatrix} =\frac{\sqrt{15}\pi}{4}\begin{pmatrix}
        0 & i & 0 \\
        i & 0 & 0\\
        0 & 0 & 0
    \end{pmatrix}
\end{align}
which is $P$ in \cite{albertini_sub-riemannian_2020} scaled by $T$. Using our formulation, the Hamiltonian is then: 
\begin{align}
    H(t) &= e^{-i\Lambda t}\sin\ad_{\Theta_*}(\Phi_*)e^{i\Lambda t}=  e^{A t}Pe^{-A t}\\
    &= i \lambda_1 \cos\omega t \pm i\lambda_6 \sin \omega t
\end{align}
for $\omega = 7/\sqrt{15}$. 

%==========DISCUSSION
\section{Discussion}
We have demonstrated that for specific categories of quantum control problems, particularly those where the antisymmetric centralizer generators parameterised by angle $\theta$ remain constant, it is possible to obtain analytic solutions for time-optimal circuit synthesis for non-exceptional symmetric spaces using a global Cartan decomposition. This is particularly true when the control subalgebras are restricted to cases where the Hamiltonian consists of generators from a horizontal distribution (bracket-generating \cite{helgason_differential_1979,knapp_representation_2001}) $\frak{p}$ with $\frak{p}\neq \frak{g}$ (where the vertical subspace is not null). Direct access is only available to subalgebras $\frak{p} \subset \frak{g}$. However, we have shown that if the assumptions $[\frak{p},\frak{p}] \subset \frak{k}$ and $d\Theta=0$ hold, arbitrary generators in $\frak{k}$ can be indirectly synthesised (via application of Lie brackets), which in turn makes the entirety of $\frak{g}$ available in optimal time. Geometrically, we have demonstrated a method for synthesis of time-optimal subRiemannian geodesics using $\frak{p}$. Note that, as mentioned in the example for $SU(2)$ above, subRiemannian geodesics may exhibit non-zero curvature with respect to the entire manifold, but this is to be expected where we are limited to a control subalgebra. Consequently, in principle, arbitrary $U_T \in G$ (rather than just) $U_T \in G/K$ becomes reachable in a control sense.

\bibliographystyle{unsrt}
\bibliography{references2}
% \printbibliography

%%==============APPENDIX============================================================%%
\newpage
\appendix
%Geodesics and algebra

\section{Generalised constant-$\theta$ method}\label{app:sec:generalmethod}
In this section, we generalise the method detailed above. Given a $G=KAK$ decomposition, an arbitrary unitary $U_T \in G$ has a decomposition as:
\begin{align}
U = ke^{i\Theta}c = qe^{ic^\inv \Theta c}
\label{eqn:generalU=keithetac}
\end{align}
where $k,c \in K$ and $e^{i\Theta} \in A$. Define Cartan conjugation as:
\begin{align}
k^\chi = k^\inv
\qquad
(e^{i\Theta})^\chi = e^{i\Theta}
\qquad
(UV)^\chi = V^\chi U^\chi
\end{align}
The Cartan projection is:
\begin{align}
\pi(U) = U^\chi U
\end{align}
which $U$ to an element of the subspace of $G$ that is fixed by $\chi$. Combining with 
equation (\ref{eqn:generalU=keithetac}) we have:
\begin{align}
\pi(ke^{i\Theta}c) = e^{2c^\inv i\Theta c},
\end{align}
That is, the representation of equation (\ref{eqn:generalU=keithetac}) as projected into the symmetric space $G/K$ (for which $\frak{k},\frak{p}$ are fixed points) establishing the symmetric space as a section of the $K$-bundle:
\begin{align}
\pi(G) \cong G/K.
\end{align}
The existence of $\chi$ and $\pi$ are sufficient for $G/K$ to be considered globally symmetric i.e. it has an involutive symmetry at every point.  The compactness of $G$ refers to the property that $G$ is a compact Lie group, meaning it is a closed and bounded subset of the Euclidean space in which it is embedded. The symmetric space $G/K$ is equipped with a has a Riemannian metric, which is a smoothly varying positive-definite quadratic form on the tangent spaces of the symmetric space. Noting the Euler formula:
\begin{align}
e^{i\Theta}Xe^{-i\Theta} = \text{Ad}_{e^{i\Theta}}(X)= e^{i\ad_\Theta}(X) = \cos\ad_\Theta(X)+i\sin\ad_\Theta(X)
\end{align}
where $X \in \frak{g}, e^{i\Theta} \in A \subset G$. Note here the (lower-case) adjoint action $\ad_\Theta(X)$ is the action of the Lie algebra generators on themselves, thus takes the form of the Lie derivative (commutator): 
\begin{align}
    \ad_\Theta(X) = [\Theta,X]
\end{align}
whereas the (upper-case) group adjoint action $\text{Ad}_{\Theta}$ is one of conjugation of group elements (hence we exponentiate the generator $X$ implicitly). The Maurer-Cartan form becomes in general:
\begin{align}
dUU^\inv
&= k\left(k^\inv dk + \cos\ad_\Theta(dcc^\inv) + id\Theta + i \sin\ad_\Theta(dcc^\inv)\right)k^\inv
\label{eqn:general:maurercartan}
\end{align}
% Note that in general:
% \begin{align}
%     d\Theta = \sum_k \partial_{\theta^k} H_k d\theta_k
% \end{align}
% where $H_k \in \frak{a}$.  
Recalling that Cartan conjugation is the negative of the relevant involution $\iota$, define the control space subalgebra as:
\begin{align}\label{controls}
\p = \{-iH : \chi(-iH) = -iH\}
\end{align}
which satisfies $\k = [\p,\p]$ and the Cartan commutation relations more generally. By restricting the Maurer-Cartan form (Hamiltonian) to the antisymmetric control subalgebra:
\begin{align}\label{constraint2}
dUU^\inv \in \p
\end{align}
we thereby define a minimal connection (see Appendix section \ref{appendix:minimalconnections}):
\begin{align}
k^\inv dk = - \cos\ad_\Theta(dcc^\inv) 
\label{eqn:general:minconnkinvdk=-cosadthetadccinv}
\end{align}
which can as per the examples be written in its parametrised form as:
\begin{align}
\dot{\psi}^{\alpha,r}+\dot{\phi}^{\alpha,r}\cos\alpha(\Theta)=0
    \label{eqn:general:connection1}
\end{align}
here $\alpha$ is a functional on $\Theta$ that selects out the relevant parameter, e.g. when $\Theta = \sum_k \theta^k H_k$ then $\alpha$ selects out the appropriate $\theta^k \in \Complex$. 

Note that if $a \in \Theta$ comprises multiple $H_k$, then  
The related Hamiltonian  may also be expressed as:
\begin{align}
H = i\frac{dU}{dt}U^\inv = -k\Big(d\Theta +  \sin\ad_\Theta(dcc^\inv) \Big)k^\inv.
\label{eqn:general:H=dtheta+sinadthetadcccinv}
\end{align}\\
Given $\g$ as the Lie algebra of $G$, define the Killing form as:
\begin{align}
(X,Y)=\frac{1}{2}\text{Re}\Tr(\ad_X\ad_Y)
\end{align}
where $\ad_X$ is the adjoint representation of $X$. The Killing form is used to define an inner product on $\frak{g}$ allowing measurement of lengths and angles in $\frak{g}$.
Define the inner product for weights and the Weyl vector $\rho$:
\begin{align}
(\alpha,\beta) = g^{kl} H_k H_l
\hspace{50pt}
\text{and}
\hspace{50pt}
\rho = \frac{1}{2}\sum_{\alpha\in R^+} \alpha = \sum_{k=1}^r \phi^k
\end{align}
where $\{H_k\}$ is a basis for the Cartan subalgebra $\frak{a}$, $R^+$ is the set of positive roots $\alpha$, $r$ is the rank, and $\{\phi^k\}$ are the fundamental weights. The weights $(\alpha,\beta)$ of a representation are the eigenvalues of the Cartan subalgebra $\frak{a}$, which (see below) is a maximal abelian subalgebra of $\frak{g}$. The Weyl vector $\rho$ is a special weight that is associated with the root system of $\frak{g}$. It is defined as half the sum of the positive roots $\alpha$ (counted with multiplicities). The Weyl vector is used in Weyl's character formula, which gives the character of a finite-dimensional representation in terms of its highest weight \cite{hall_lie_2013} which we denote below as $\tau$. \\
\\
Define the Euclidean norm using the Killing form as:
\begin{align}
|X| = \sqrt{(X,X)}
\end{align}
We leverage the fact that the Killing form is quadratic for semi-simple Lie algebras $\frak{g}$ such that:
\begin{align}
|idUU^\inv|^2 = |ik^\inv dk + \cos\ad_\Theta(idcc^\inv)|^2 + |d\Theta|^2 + |\sin\ad_\Theta(dcc^\inv)|^2
\label{eqn:general-maurer-cartainsquared}
\end{align}
Let the Hamiltonian have an isotropic cutoff:
\begin{align}
|H|=\Omega.
\end{align}
By the Schrodinger equation, the time elapsed over the path $\gamma$ is given by:
\begin{align}
\Omega t = \Omega \int_\gamma dt = \int_\gamma |idUU^\inv| =  \int_\gamma \left|i\frac{dU}{ds}U^\inv\right| ds.
\end{align}
Define:
\begin{align}
\dot t = \left|i\frac{dU}{ds}U^\inv\right|.
\end{align}
with:
\begin{align}
T =  \min_\gamma t.
\end{align}
To perform the local minimization, we introduce a vector of Lagrange multipliers (defined below in equation (\ref{eqn:general:lagrangevector})):
\begin{align}
\Omega t =  \int_\gamma\Big(\dot{t} + \big(\Lambda,k^\inv \dot{k} + \cos\ad_\Theta(\dot{c}c^\inv)\big)\Big) ds
\label{eqn:general:Omegat=intgamma(...)}
\end{align}
note
\begin{align}
k\left(k^\inv \dot{k} + \cos\ad_\Theta(\dot{c}c^\inv)\right)k^\inv = \frac{\dot{U}U^\inv -\chi(\dot{U}U^\inv)}{2}
\end{align}
and
\begin{align}
k\left(i\dot\Theta + i \sin\ad_\Theta(\dot{c}c^\inv)\right)k^\inv = \frac{\dot{U}U^\inv +\chi(\dot{U}U^\inv)}{2}
\end{align}
We can further simplify via expanding in the restricted Cartan-Weyl basis, noting that $\alpha$ indexes the relevant roots and $s$ indexes the relevant sets of roots $r \in \Delta$). The restricted Cartan-Weyl basis allows $\frak{g}$ to be decomposed as the sum of the commutant basis vectors $H_k \in \frak{a}$ and the root vectors $E_\alpha$:
\begin{align}
    \frak{g} = H_k \bigoplus E_\alpha
\end{align}
where there are $r$ such positive roots. The Maurer-Cartan form becomes (equation (\ref{eqn:general:Omegat=intgamma(...)})) then becomes expressed in terms of weights and weight vectors:
\begin{align}
\Theta = H_k \theta^k,
\label{eqn:general:theta=hkthetak}
\end{align}
\begin{align}
i\dot{c}c^\inv = F_{\alpha,r}\dot{\phi}^{\alpha,r}
\label{eqn:general:idotccinv=Ialphadotphialphas}
\end{align}
\begin{align}
ik^\inv\dot{k} = F_{\alpha,r}\dot{\psi}^{\alpha,r}+\frak{m}
\label{eqn:general:ikinvkdot=Ialphaspsi+a}
\end{align}
and
\begin{align}
\Lambda = F_{\alpha,r}\lambda^{\alpha,r}+\frak{m}
\label{eqn:general:lagrangevector}
\end{align}
In the above equations, the $\alpha$ are summed over restricted positive roots (of which there are $r$ many). $F_\alpha \in \k - \m$  with $\dot\phi,\dot\psi \in \Complex$ coefficients. The Lagrange multiplier vector $\Lambda$ in equation (\ref{eqn:general:lagrangevector}) is in generalised form. Note that in the $SU(2)$ case above, the simplicity of $\frak{k} = \{ J_z \}$ simplifies the multiplier term in the action equation (\ref{eqn:su2:s=omegataction}). Here the Cartan subalgebra $\h$ is given by:
\begin{align}
    \h = \{\a,\m\}
\end{align}
This algebra is distinct from a maximally compact Cartan algebra in that $\h$ intersects with both $\p$ (the non-compact part of $\h$) and $\k$ (the compact part of $\h$). 
In many cases, the elements of $\h$ are themselves diagonal (see for example Hall and others \cite{hall_lie_2013,cahn_semi-simple_2014}) and entirely within $\k$. In our case, $\a \in \p$ but $\m \in k$. The construction of the restricted Cartan-Weyl basis is via the adjoint action of $\a$ on the Lie algebra, such that it gives rise to pairings $F_\alpha \in \k-\m, E_\alpha \in \p-\a$ conjugate under the adjoint action. 

The Cartan-Weyl basis has the property that the commutation relations between the basis elements are simplified because (i) the Cartan generators (belonging to an abelian subalgebra) commute and (ii) the commutation relations between a Cartan generator and a root vector are proportional to the root vector itself. The commutation relations between two root vectors can be more complicated, but they are determined by the structure of the root system. The set of roots, which are the non-zero weights of the adjoint representation of the Lie algebra. The roots form a discrete subset of the dual space to the Cartan subalgebra, and they satisfy certain symmetries and relations that are encoded in the Dynkin diagram of the Lie algebra \cite{knapp_representation_2001,helgason_differential_1979}.Transforming to the restricted Cartan-Weyl basis allows us to represent the parametrised form of equation (\ref{eqn:general-maurer-cartainsquared}) as:
\begin{align}
|\dot\Theta|^2 = g_{kl}\dot\theta^k\dot\theta^l,
\end{align}
\begin{align}
| \sin\ad_\Theta(\dot{c} c^\inv)|^2 = \sum_{\alpha,r}g_{\alpha\alpha} (\dot{\phi}^{\alpha,r})^2\sin^2\alpha(\Theta),
\end{align}
\begin{align}
|k^\inv \dot{k} + \cos\ad_\Theta(\dot{c}c^\inv)|^2 = \sum_{\alpha,r}g_{\alpha\alpha} \big(\dot{\psi}^{\alpha,r}+\dot{\phi}^{\alpha,r}\cos\alpha(\Theta)\big)^2 + |\frak{m}|^2,
\end{align}
and using the vector of Lagrange multipliers with the connection:
\begin{align}
\big(\Lambda,k^\inv \dot{k} + \cos\ad_\Theta(\dot{c}c^\inv)\big)
= \sum_{\alpha,r}g_{\alpha\alpha} \lambda^{\alpha,r}\big(\dot{\psi}^{\alpha,r}+\dot{\phi}^{\alpha,r}\cos\alpha(\Theta)\big) + |\frak{m}|^2
\end{align}
Here we have used the fact that the only non-vanishing elements of the Killing form are $g_{\alpha \alpha}$ and $g_{jk}$ (and using $g_{\alpha\alpha} = (E_\alpha,E_\alpha)$), the inner product of the basis elements with themselves, and the restricted Cartan-Weyl basis to simplify the variational equations. The nonzero functional derivatives are:
\begin{align}
\Omega \frac{\delta t}{\delta \dot{\psi}^{\alpha,r}} = g_{\alpha\alpha}\left(\frac{1}{\dot{t}}\big(\dot\psi^{\alpha,r}+\dot\phi^{\alpha,r}\cos\alpha(\Theta)\big) + \lambda^{\alpha,r}\right)
\label{eqn:general:ELGdotphi}
\end{align}
\begin{align}
\Omega \frac{\delta t}{\delta \dot{\phi}} = g_{\alpha\alpha}\left(\frac{\dot\phi^{\alpha,r}}{\dot{t}} + \left(\frac{\dot\psi^{\alpha,r}}{\dot{t}}+\lambda^{\alpha,r}\right)\cos\alpha(\Theta)\right)
\end{align}
\begin{align}
\Omega \frac{\delta t}{\delta \dot{\theta}^k} = g_{kl}\frac{\dot\theta^l}{\dot{t}}
\end{align}
\begin{align}
\Omega \frac{\delta t}{\delta \theta^k}
= -\sum_{\alpha,r}g_{\alpha\alpha}\left(\frac{\dot\psi^{\alpha,r}}{\dot{t}}+\lambda^{\alpha,r}\right)\dot\phi^{\alpha,r}\alpha(H_k)\sin\alpha(\Theta)
\end{align}
\begin{align}
\Omega \frac{\delta t}{\delta \lambda^{\alpha,r}}
=\dot\psi^{\alpha,r}+\dot\phi^{\alpha,r}\cos\alpha(\theta)
\label{eqn:general:ELGlambda}
\end{align}
From the Euler-Lagrange equations and by design from the connection we have the constraints:
\begin{align}
k^\inv \dot{k} =- \cos\ad_\Theta(\dot{c}c^\inv)
\end{align}
We assume again the Lagrange multipliers are constant. As for the case of $SU(2)$, each Lagrange multiplier $\lambda^{\alpha,r}$ then becomes a global gauge degree of freedom in the sense that:
\begin{align}
\frac{\partial T}{\partial\lambda^{\alpha,r}}=0.
\end{align}
with $\dot\psi(s)$ becoming local gauge degrees of freedom  under the constraint:
\begin{align}
\frac{\delta T}{\delta \dot\psi^{\alpha,r}}=0.
\end{align}
We are free to choose the gauge trajectory $\dot\psi(s)$ as:
\begin{align}
k^\inv \dot{k}/\dot{t}+\Lambda=0
\end{align}
With this choice, we have also by the remaining Euler-Lagrange equations:
\begin{align}
\dot{c}c^\inv/\dot{t}=\text{constant}
\end{align}
and
\begin{align}
\dot\Theta/\dot{t}=\text{constant}.
\end{align}
Minimizing over the constant $\dot\theta$:
\begin{align}
\Omega \frac{\partial T}{\partial \dot\theta^k}=g_{kl}(\dot\theta^l/\dot{t})\int_\gamma ds = 0,
\end{align}
we see
\begin{align}
\Theta = \text{constant}
\end{align}
The above functional equations show that when the action is varied, each term in equation (\ref{eqn:general:maurercartan}) vanishes apart from the $\sin \text{ad}_\Theta(dcc^\inv)$ term. 
Calculating optimal evolution time then reduces to minimization over initial conditions:
\begin{align}
T =  \min_{\Theta,\Phi} \left|\sin\ad_\Theta(\Phi)\right|
\label{eqn:general:Tsinadthetaphi}
\end{align}
where
\begin{align}
\Phi = \int_\gamma -i dc c^\inv. 
\end{align}
We can express in terms of the Cartan-Weyl basis as follows:
\begin{align}
    \Phi&= \sum_{\alpha,r} \dot\phi^{\alpha,r} F_{\alpha,r} + \frak{m}
    \label{eqn:general:PhisumphiFalpharplusm}
\end{align}
where: 
\begin{align}
    &F_{\alpha,r} =\frac{1}{\alpha(\Theta)}\ad_\Theta (E_{\alpha,r}) \in \frak{k}-\frak{m} \label{eqn:general:Falphar}\\
    &\ad^2_\Theta (E_{\alpha,r}) = \alpha(\Theta)^2 E_{\alpha,r} \in \frak{p} - \frak{a} \label{eqn:general:ad2Ealphr}\\
    &\ad_\Theta(\frak{m}) = 0 \label{eqn:general:adthetam}
\end{align}
such that:
\begin{align}
    \sin\ad_\Theta(\dot\Phi) = \sum_{\alpha,r}\dot\phi^{\alpha,r}\sin\alpha(\Theta)E_{\alpha,r} \in \frak{p}
    \label{eqn:general:sinadthetaPhiEalphar}
\end{align}
The minimisation can be expressed in terms of parameters noting the trace as a rank-2 tensor contraction. The problem of finding the minimal time Hamiltonian is thereby simplified considerably. Consider targets of the form:
\begin{align}
U(T)U_0^\inv = e^{-iX} \in K
\end{align}
Again:
\begin{align}
    U(T)U_0^\inv= e^{-iX} =   ke^{i\Theta}c = qe^{ic^\inv \Theta c}
\end{align}
Explicitly we equate:
\begin{align}
    U(T) = q \qquad U(0) = e^{ic^\inv \Theta c} 
\end{align}
As was the case for $SU(2)$, we select initial conditions as:
\begin{align}
U_0 = e^{i\Theta}
\end{align}
where we can see a quantization condition emerges by satisfying the commutant condition that $c = \exp(\Phi) \in K$ (with $\Phi \in \k$) commute with $\Theta$:
\begin{align}
\Big\{\exp(\Phi) \in K : \exp(\Phi)\Theta \exp(-\Phi) =\Theta \Big\}  = M
\label{eqn:general:csubsetidentity}
\end{align}
Such a condition is equivalent to $\Theta c \Theta^\inv = c$. In the SU(2) case, because we only have a single generator in $\Phi$, the condition manifests in requiring the group elements $c \in K$ to resolve to $\pm \mathbb{I}$, which in turn imposes a requirement that their parameters $\phi_j$ be of the form $\phi_j = 2\pi n$ for $n \in \mathbb{Z}$ in order for $e^{i\phi_j k_j } = \mathbb{I}$ where $k \in \frak{k}$. In general, however, the commutant will have a nontrivial connected subgroup and it still ``quantizes'' into multiple connected components. In practice this means that $\phi_k$ are not, in general, integer multiples of $2\pi$ and instead must be chosen to meet the commutant condition in each case. We note that:
\begin{align}
q(T)  = k(T)c(T) = e^{-iX}
\end{align}
or equivalently
\begin{align}
X &= \int_\gamma i dq q^\inv\\
& = \int_\gamma k\left(ik^\inv dk + idc c^\inv\right)k^\inv\\ 
& = \int_\gamma k\left(1-\cos\ad_\Theta \right)\big(idc c^\inv\big)k^\inv\\ 
& = -\left(1-\cos\ad_\Theta \right)\big(\Phi\big)
\end{align}
where we have used the minimal connection in equation (\ref{eqn:general:minconnkinvdk=-cosadthetadccinv}), where the last equality comes from $k(0)=\mathbb{I}$ and the arc-length parametrisation of $\gamma$ between $s\in [0,1]$. The optimal time is:
\begin{align}
\Omega T = \min_{\Theta,\Phi} \big|\sin\ad_\Theta(\Phi)\big|
\label{eqn:general:optimaltime}
\end{align}
with minimization over constraints:
\begin{align}
e^{\Phi} \in M
\hspace{50pt}
\text{and}
\hspace{50pt}
X =  \left(1-\cos\ad_\Theta \right)(\Phi)
\label{eqn:general:optimaltimeconstraints}
\end{align}
Time-optimal control is given by Hamiltonians of the form:
\begin{align}
H(t)= e^{-i\Lambda t}\sin\ad_{\Theta_*}(\Phi_*)e^{i\Lambda t}
\label{eqn:general:timeoptimalhamiltonian}
\end{align}
where $\dot\Phi = \Phi/T$, $\Theta_*$ and $\Phi_*$ are the critical values which minimize the time,
and the multiplier is the turning rate
\begin{align}
\Lambda & = -k^\inv \dot{k}/\dot t\\
& = -\int_\gamma k^\inv \dot{k} /T\\
& = \frac{\cos\ad_{\Theta_*}(\Phi_*)}{T}.
\label{eqn:general:lambda=cosadthetaphi/T}
\end{align}
The global minimization then depends upon the choice of Cartan subalgebra $\a \ni \Theta$ (as illustrated in the examples above).

%========ADJOINT ACTION
\newpage
\section{Adjoint action and commutation relations} \label{app:su3commutrelat}
The adjoint action of group elements can be expressed via conjugation as $\text{Ad}_h(g) = hgh^\inv$where $g,h \in G$. The adjoint action of a Lie algebra upon itself, which we focus on, is via the commutator namely $\ad_X(Y) = [X,Y]$ where $X,Y \in \frak{g}$. Thus terms such as $\ad_\Theta(\Phi)$ involve calculation of commutation relations among generators. Note the following used in the paper:
\begin{align}
    e^{i\Theta}Xe^{-i\Theta} = e^{i\ad_\Theta}(X) = \cos\ad_\Theta(X)+i\sin\ad_\Theta(X)
    \label{eqn:econjsinhcosh}
\end{align}
The $\cos \ad_\Theta (X)$ term can be understood in terms of the cosine expansion:
\begin{align}
    \cos \ad_\Theta (X) = \sum_{n=0}^\infty \frac{(-1)^n}{(2n)!} (\ad_\Theta))^{2n} (X)
\label{eqn:cosadthetaXexpansion}
\end{align}
Thus each term is a multiple of $\ad_\theta^2$. For [certain] choices of generators $\Theta$ and $X \in \frak{k}$ this effectively means that [acts as an involution up to coefficients]. Thus in the $SU(2)$ case, $\ad^{2n}_{J_y}(J_z) \propto J_z$ (with $n=0,1,2,...$). Each application of the adjoint action acquires a $\theta^2$ term (from $\Theta = i\theta J_y$), such that the series can be written:
\begin{align}
    \cos \ad_{\Theta} (-iJ_z) = \sum_{n=0}^\infty \frac{(-1)^n}{(2n)!} (\theta))^{2n} (J_z) = \cos(\theta)(-iJ_z)
\label{eqn:su2cosadthetaXexpansion}
\end{align}
More generally, the Cartan commutation relations exhibit such an [involutive] property given that for $\Theta \in \frak{k}$:
\begin{align} \ad_\Theta(\frak{k})=[\Theta,\frak{k}] \in \frak{p} \qquad \ad^2_ \Theta(\frak{k})=[\Theta,[\Theta,\frak{k}]] \in \frak{k}
\end{align}
In the general case, assuming an appropriately chosen representation, each application of the adjoint action by even $n$ or odd $2n+1$ powers (and thus our $\cos \theta$ and $\sin \theta$ terms) will be scaled by an eigenvalue $\alpha$, such eigenvalue being the [root] (see below for an example). 
Thus we have $\ad_{\Theta}^{2n}(\frak{k}) \subset \frak{k}$ and:
\begin{align}
\cos(\ad_\Theta) \Phi = \cos\alpha(\theta)\Phi 
\label{eqn:cosadthetaphi=costhetaphi}
\end{align}
such as in equation (\ref{eqn:main-x=1-cosadthetabigphi}) and similarly for the sine terms. For the $\sin\ad_\Theta(X)$ terms, by contrast, the adjoint action is odd: 
\begin{align}
    \sin \ad_\Theta (X) = \sum_{n=0}^\infty \frac{(-1)^{n}}{(2n+1)!} (\ad_\Theta))^{2n+1} (X)
\label{eqn:sinadthetaXexpansion}
\end{align}
such that $\ad_{\Theta}^{2n+1}(\frak{k}) \in \frak{p}$, hence the sine term arising in our Hamiltonian form equation (\ref{eqn:general:timeoptimalhamiltonian}) which, when conjugated with $\Lambda$ results in a Hamiltonian given in terms of control generators, namely:
\begin{align}
    [\sin \ad_\Theta(\frak{k}),\Lambda] \subset \frak{p}
\end{align}

\section{Minimal connections and holonomy groups}\label{appendix:minimalconnections}
A simple application of the product rule and the Euler formula:
\begin{align}
e^{i\Theta}Xe^{-i\Theta} = e^{i\ad_\Theta}(X) = \cos\ad_\Theta(X)+i\sin\ad_\Theta(X)
\end{align}
reveals:
\begin{align}
dU U^\inv = k\left(k^\inv dk + \cos\ad_\Theta(dc c^\inv)+id\Theta + i \sin\ad_\Theta(dc c^\inv)\right)k^\inv.
\end{align}
Symmetric controls:
\begin{align}
-iHdt \in \p
\end{align}
which satisfy the Lie triple property:
\begin{align}
[\p,[\p,\p]]\subset\p
\end{align}
define the connection:
\begin{align}
k^\inv dk = - \cos\ad_\Theta(dc c^\inv)
\end{align}
which is minimal in the sense that it minimizes the invariant line element
\begin{align}
\min_{k\in K}{|idU U^\inv|^2} =\min_{k\in K}{|ik^\inv dk + i\cos\ad_\Theta(dc c^\inv)|^2} + |d\Theta|^2+| \sin\ad_\Theta(dc c^\inv)|^2
\end{align}
where the minimization is over curves, $k(t)$.
This means that minimization with a horizontal constraint is equivalent to minimization over the entire isometry space.
Such submanifolds are said to be totally geodesic. Targets $U(T) = e^{-iX}$ with effective Hamiltonians:
\begin{align}
-iX \in \k \equiv [\p,\p]
\end{align} 
are known as holonomies because they are generators of \textit{holonomic groups}, groups whose action on a representation (e.g. the relevant vector space) causes vectors to be parallel-transported in space i.e. the covariant derivative vanishes along such curves traced out by the group action. Intuitively, one can consider holonomic groups as orbits such that any transformation generated by elements of $\frak{k}$ will `transport' the relevant vector along paths represented as geodesics. However, such vectors are constrained to such orbits if those generators are only drawn from $\frak{k}$ i.e. $[k,k] \subseteq k$ for a chosen subalgebra $\frak{k}$. To transport vectors elsewhere in the space, one must apply generators in $\frak{p}$ which is analogous to shifting a vector to a new orbit.  Although in application, one considers $U(0)=1$, it is important to remember this variational problem is right-invariant, so one could just as well let $U(0)=U_0$ be arbitrary, as long as the target is understood to correspondingly be $U(T) = e^{-iX}U_0$.

\end{document}